\documentclass[twocolumn]{aastex63}
\pdfoutput=1
\usepackage{amsmath}
\usepackage{amsfonts}
\usepackage{amssymb}
\usepackage{chngcntr}

\usepackage{graphicx}
\usepackage[varg]{txfonts}
\usepackage{mwe} 
\usepackage{url}
\usepackage{longtable}
\usepackage{float}

\usepackage{makecell}
\usepackage{xspace}
\usepackage{xcolor}

\usepackage{scrextend}
\usepackage{hyperref}

\newcommand{\fulltarget}{SDSS J162829.17+432948.5}
\newcommand{\target}{J1628+4329}

\newcommand{\A}{\ensuremath{\mathrm{\AA}}\xspace}

\newcommand{\Hg}{\ensuremath{\rm{H}\gamma}\xspace}
\newcommand{\Hb}{\ensuremath{\rm{H}\beta}\xspace}
\newcommand{\Ha}{\ensuremath{\rm{H}\alpha}\xspace}
\newcommand{\OIII}{[\ion{O}{iii}]\ensuremath{\,\lambda5007}\xspace}

\newcommand{\tdyn}{\ensuremath{\tau_{\rm{dyn}}}\xspace}
\newcommand{\Rcloud}{\ensuremath{R_{\rm{cloud}}}\xspace}
\newcommand{\Msunyr}{\ensuremath{M_\odot/\rm{yr}}\xspace}

\newcommand{\Lop}{\ensuremath{L_{5100}}}

\newcommand{\NLCO}{three}
\newcommand{\los}{LOS}

\newcommand{\todo}{\ifmmode \text{\color{red}{\Huge(\bullet)}} \else {\color{red}{\Huge$\bullet$}}\fi}
\newcommand{\tido}{\ifmmode {{\color{red}\bullet}} \else \color{red}{$\bullet$}\fi}

\newcommand{\E        }[1]{\ifmmode 10^{#1} \else $10^{#1}$\fi}
\newcommand{\tE        }[1]{\ifmmode \times10^{#1} \else $\times10^{#1}$\fi}
\newcommand{\til}{\ifmmode \sim \else $\sim$\fi}
\newcommand{\pc}	{\ifmmode {\rm pc} \else pc\fi}
\newcommand{\kpc}	{\ifmmode {\rm kpc} \else kpc\fi}
\newcommand{\ld}	{\ifmmode {\rm lt-day} \else lt-day\fi}
\newcommand{\kms}	{\ifmmode {\rm km\,s}^{-1} \else km\,s$^{-1}$\fi}
\newcommand{\cc}	{\ifmmode {\rm cm}^{-3}    \else cm$^{-3}$\fi}
\newcommand{\cmii}	{\ifmmode {\rm cm}^{-2}    \else cm$^{-2}$\fi}
\newcommand{\ergs}	{\ifmmode {\rm erg\,s}^{-1} \else erg s$^{-1}$\fi}
\newcommand{\ergcms}	{\ifmmode {\rm erg\,cm}^{-2}\,{\rm s}^{-1} \else erg\,cm$^{-2}$\,s$^{-1}$\fi}
\newcommand{\ergcmsA}	{\ifmmode {\rm erg\,cm}^{-2}\,{\rm s}^{-1}\,{\rm\AA}^{-1}
\else erg\,cm$^{-2}$\,s$^{-1}$\,\AA$^{-1}$\fi}
\newcommand{  \ergcmsHz  }{\ifmmode{\rm erg\,cm}^{-2}\,{\rm s}^{-1}\,{\rm Hz}^{-1}
                       \else ergs\,cm$^{-2}$\,s$^{-1}$\,Hz$^{-1}$\fi}
\newcommand{\kev}	{\ifmmode {\rm keV} \else keV\fi}

\newcommand{\mic}	{\ifmmode {\rm \mu m} \else $\mu$m\fi}
\newcommand{\vFWHM}	{\ifmmode v_{\mbox{\tiny FWHM}} \else $v_{\mbox{\tiny FWHM}}$\fi}
\newcommand{\vBLR}	{\ifmmode v_{\mbox{\tiny BLR}} \else $v_{\mbox{\tiny BLR}}$\fi}
\newcommand{\sigBLR}	{\ifmmode \sigma_{\mbox{\tiny BLR}} \else $\sigma_{\mbox{\tiny BLR}}$\fi}
\newcommand{\vNLR}	{\ifmmode v_{\mbox{\tiny NLR}} \else $v_{\mbox{\tiny NLR}}$\fi}
\newcommand{\tauBLR}	{\ifmmode \tau_{\mbox{\tiny BLR}} \else $\tau_{\mbox{\tiny BLR}}$\fi}

\newcommand{\Hubble}	{\ifmmode {\rm km\,s}^{-1}\,{\rm Mpc}^{-1} \else km\,s$^{-1}$\,Mpc$^{-1}$\fi}
\newcommand{\NDunit}	{\ifmmode {\rm Mpc}^{-3} \else Mpc$^{-3}$\fi}
\newcommand{\LFunit}	{\ifmmode {\rm Mpc}^{-3}\,{\rm mag}^{-1} \else Mpc$^{-3}$\,mag$^{-1}$\fi}
\newcommand{\MFunit}	{\ifmmode {\rm Mpc}^{-3}\,{\rm dex}^{-1} \else Mpc$^{-3}$\,dex$^{-1}$\fi}

\newcommand{\Msun}{\ifmmode M_{\odot} \else $M_{\odot}$\fi}
\newcommand{\Lsun}{\ifmmode L_{\odot} \else $L_{\odot}$\fi}
\newcommand{\Zsun}{\ifmmode Z_{\odot} \else $Z_{\odot}$\fi}
\newcommand{\mpyr}{\ifmmode \Msun\,{\rm yr}^{-1} \else $\Msun\,{\rm yr}^{-1}$\fi}

\newcommand{\qnote}{\ifmmode q_{0} \else $q_{0}$\fi}
\newcommand{\Hnote}{\ifmmode H_{0} \else $H_{0}$\fi}
\newcommand{\hnote}{\ifmmode h_{0} \else $h_{0}$\fi}
\newcommand{\anote}{\ifmmode a_{0} \else $a_{0}$\fi}
\newcommand{\tnote}{\ifmmode t_{0} \else $t_{0}$\fi}

\newcommand{  \Halpha   }{\ifmmode {\rm H}\alpha \else H$\alpha$\fi}

\newcommand{  \ha       }{\Halpha}
\newcommand{  \Hbeta    }{\ifmmode {\rm H}\beta \else H$\beta$\fi}

\newcommand{  \hb       }{\Hbeta}
\newcommand{  \Hgamma   }{\ifmmode {\rm H}\gamma \else H$\gamma$\fi}
\newcommand{  \Hdelta   }{\ifmmode {\rm H}\delta \else H$\delta$\fi}
\newcommand{  \Lya      }{\ifmmode {\rm Ly}\alpha \else Ly$\alpha$\fi}
\newcommand{  \Lyb      }{\ifmmode {\rm Ly}\beta \else Ly$\beta$\fi}
\newcommand{  \Pa       }{\ifmmode {\rm P}\alpha \else P$\alpha$\fi}
\newcommand{  \Pb       }{\ifmmode {\rm P}\beta \else P$\beta$\fi}
\newcommand{  \Bra      }{\ifmmode {\rm Br}\alpha \else Br$\alpha$\fi}
\newcommand{  \Brg      }{\ifmmode {\rm Br}\gamma \else Br$\gamma$\fi}
\newcommand{  \hii      }{\ifmmode {\rm H}\,\textsc{ii} \else H\,\textsc{ii}\fi}
\newcommand{  \hei      }{\ifmmode {\rm He}\,\textsc{i} \else He\,\textsc{i}\fi}
\newcommand{  \heii     }{\ifmmode {\rm He}\,\textsc{ii} \else He\,\textsc{ii}\fi}
\newcommand{  \HeIIuv   }{\ifmmode {\rm He}\,\textsc{ii}\,\lambda1640 \else He\,\textsc{ii}\,$\lambda1640$\fi}
\newcommand{  \HeIIop   }{\ifmmode {\rm He}\,\textsc{ii}\,\lambda4686 \else He\,\textsc{ii}\,$\lambda4686$\fi}
\newcommand{  \CII	}{\ifmmode \left[{\rm C}\,\textsc{ii}\right]\,\lambda157.74\,\mu{\rm m} \else [C\,{\sc ii}]\ $\lambda157.74\,\mu{\rm m}$\fi}
\newcommand{  \cii	}{\ifmmode \left[{\rm C}\,\textsc{ii}\right] \else [C\,{\sc ii}]\fi}

\newcommand{  \ciii     }{\ifmmode {\rm C}\,\textsc{iii}\right] \else C\,\textsc{iii}]\fi}
\newcommand{  \CIII     }{\ifmmode {\rm C}\,\textsc{iii}\right]\,\lambda1909 \else C\,\textsc{iii}]\,$\lambda1909$\fi}
\newcommand{  \civ      }{\ifmmode {\rm C}\,\textsc{iv}  \else C\,\textsc{iv}\fi}
\newcommand{  \CIV      }{\ifmmode {\rm C}\,\textsc{iv}\,\lambda1549 \else C\,\textsc{iv}\,$\lambda1549$\fi}
\newcommand{  \NIIopt   }{\ifmmode \left[{\rm N}\,\textsc{ii}\right]\,\lambda6584 \else [N\,\textsc{ii}]\,$\lambda6584$\fi}
\newcommand{  \nii      }{\ifmmode \left[{\rm N}\,\textsc{ii}\right]  \else [N\,\textsc{ii}]\fi}
\newcommand{  \niii     }{\ifmmode {\rm N}\,\textsc{iii} \else N\,\textsc{iii}\fi}
\newcommand{  \NIII     }{\ifmmode {\rm N}\,\textsc{iii}\,\lambda4640 \else N\,\textsc{iii}\,$\lambda4640$\fi}
\newcommand{  \niv      }{\ifmmode {\rm N}\,\textsc{iv}  \else N\,\textsc{iv}\fi}
\newcommand{  \NIVuv    }{\ifmmode {\rm N}\,\textsc{iv}\,\lambda1486 \else N\,\textsc{iv}\,$\lambda1486$\fi}
\newcommand{  \nv       }{\ifmmode {\rm N}\,\textsc{v}   \else N\,\textsc{v}\fi}
\newcommand{\oi}{\ifmmode \left[{\rm O}\,\textsc{i}\right] \else [O\,{\sc i}]\fi}
\newcommand{\OI}{\ifmmode \left[{\rm O}\,\textsc{i}\right]\,\lambda6300 \else [O\,{\sc i}]$\,\lambda6300$\fi}
\newcommand{\oii}{\ifmmode \left[{\rm O}\,\textsc{ii}\right] \else [O\,{\sc ii}]\fi}
\newcommand{\OII}{\ifmmode \left[{\rm O}\,\textsc{ii}\right]\,\lambda3727 \else [O\,{\sc ii}]\,$\lambda3727$\fi}
\newcommand{\oiii}{\ifmmode \left[{\rm O}\,\textsc{iii}\right] \else [O\,{\sc iii}]\fi}
\newcommand{  \OIIIbf   }{\ifmmode {\rm O}\,\textsc{iii}\,\lambda3133 \else O\,\textsc{iii}\,$\lambda3133$\fi}
\newcommand{  \OIIIuv   }{\ifmmode {\rm O}\,\textsc{iii}\,\lambda1663 \else O\,\textsc{iii}\,$\lambda1663$\fi}
\newcommand{  \oiv      }{\ifmmode {\rm O}\,\textsc{iv}  \else O\,\textsc{iv}\fi}
\newcommand{  \OIVuv    }{\ifmmode {\rm O}\,\textsc{iv}\,\lambda1402  \else O\,\textsc{iv}\,$\lambda1402$\fi}
\newcommand{  \OIVIR    }{\ifmmode {\rm O}\,\textsc{iv}\,25.9\,\mu {\rm m} \else O\,\textsc{iv}\,$25.9\,\mu$m\fi}
\newcommand{  \ovi      }{\ifmmode {\rm O}\,\textsc{vi}   \else O\,\textsc{vi}\fi}
\newcommand{  \Ovi      }{\ifmmode {\rm O}\,\textsc{vi}\,\lambda1035 \else O\,\textsc{vi}\,$\lambda1035$\fi}
\newcommand{  \nei      }{\ifmmode {\rm Ne}\,\textsc{i}   \else Ne\,\textsc{i}\fi}
\newcommand{  \neii     }{\ifmmode {\rm Ne}\,\textsc{ii}  \else Ne\,\textsc{ii}\fi}
\newcommand{  \NeiiIR   }{\ifmmode {\rm Ne}\,\textsc{ii}\,12.8\,\mu {\rm m} \else Ne\,\textsc{ii}\,$12.8\,\mu$m\fi}
\newcommand{  \neiii    }{\ifmmode {\rm Ne}\,\textsc{iii} \else Ne\,\textsc{iii}\fi}
\newcommand{  \neiv     }{\ifmmode {\rm Ne}\,\textsc{iv}  \else Ne\,\textsc{iv}\fi}
\newcommand{  \nev      }{\ifmmode {\rm Ne}\,\textsc{v}   \else Ne\,\textsc{v}\fi}
\newcommand{  \NevIR    }{\ifmmode {\rm Ne}\,\textsc{v}\,24.3\,\mu {\rm m} \else Ne\,\textsc{v}\,$24.3\,\mu$m\fi}
\newcommand{  \nevi     }{\ifmmode {\rm Ne}\,\textsc{vi}  \else Ne\,\textsc{vi}\fi}
\newcommand{  \mgi      }{\ifmmode {\rm Mg}\,\textsc{i} \else Mg\,\textsc{i}\fi}
\newcommand{  \mgii     }{\ifmmode {\rm Mg}\,\textsc{ii} \else Mg\,\textsc{ii}\fi}
\newcommand{  \MgII     }{\ifmmode {\rm Mg}\,\textsc{ii}\,\lambda2798 \else Mg\,\textsc{ii}\,$\lambda2798$\fi}
\newcommand{  \sii      }{\ifmmode {\rm S}\,\textsc{ii} \else S\,\textsc{ii}\fi}
\newcommand{  \siii     }{\ifmmode {\rm S}\,\textsc{iii} \else S\,\textsc{iii}\fi}
\newcommand{  \siv      }{\ifmmode {\rm S}\,\textsc{iv} \else S\,\textsc{iv}\fi}
\newcommand{  \sili     }{\ifmmode {\rm Si}\,\textsc{i}   \else Si\,\textsc{i}\fi}
\newcommand{  \silii    }{\ifmmode {\rm Si}\,\textsc{ii}  \else Si\,\textsc{ii}\fi}
\newcommand{  \Siliv    }{\ifmmode {\rm Si}\,\textsc{iv}  \else Si\,\textsc{iv}\fi}
\newcommand{  \SilIVuv  }{\ifmmode {\rm Si}\,\textsc{iv}\,\lambda1400  \else Si\,\textsc{iv}\,$\lambda1400$\fi}
\newcommand{  \AlIII   }{\ifmmode {\rm Al}\,\textsc{iii}\,\lambda1857 \else Al\,\textsc{iii}\,$\lambda1857$\fi}
\newcommand{  \Aliii   }{\ifmmode {\rm Al}\,\textsc{iii} \else Al\,\textsc{iii}\fi}
\newcommand{  \caii     }{\ifmmode {\rm Ca}\,\textsc{ii} \else Ca\,\textsc{ii}\fi}
\newcommand{  \feii     }{\ifmmode {\rm Fe}\,\textsc{ii} \else Fe\,\textsc{ii}\fi}
\newcommand{  \feiii    }{\ifmmode {\rm Fe}\,\textsc{iii} \else Fe\,\textsc{iii}\fi}
\newcommand{  \Kalpha   }{\ifmmode {\rm K}\alpha \else K$\alpha$\fi}

\newcommand{ \Lhb   }{\ifmmode L_{\hb} \else $L_{\hb}$\fi}
\newcommand{ \Lha   }{\ifmmode L_{\ha} \else $L_{\ha}$\fi}
\newcommand{ \fwhb  }{\ifmmode {\rm FWHM}\left(\hb\right) \else FWHM(\hb)\fi}
\newcommand{\sighb  }{\ifmmode \sigma\left(\hb\right) \else $\sigma\left(\hb\right)$\fi}
\newcommand{ \ewhb  }{\ifmmode {\rm EW}\left(\hb\right) \else EW(\hb)\fi}
\newcommand{ \fwha  }{\ifmmode {\rm FWHM}\left(\ha\right) \else FWHM(\ha)\fi}
\newcommand{ \ewha  }{\ifmmode {\rm EW}\left(\ha\right) \else EW(\ha)\fi}
\newcommand{ \Lmg   }{\ifmmode L\left(\mgii\right) \else $L\left(\mgii\right)$\fi}
\newcommand{ \fwmg  }{\ifmmode {\rm FWHM}\left(\mgii\right) \else FWHM(\mgii)\fi}
\newcommand{ \Lciv  }{\ifmmode L\left(\civ\right) \else $L\left(\civ\right)$\fi}
\newcommand{ \fwciv }{\ifmmode {\rm FWHM}\left(\civ\right) \else FWHM(\civ)\fi}
\newcommand{ \fwhm  }{\ifmmode {\rm FWHM} \else FWHM\fi} 
\newcommand{ \voff  }{\ifmmode v_{\rm off} \else $v_{\rm off}$\fi} 
\newcommand{ \vmax  }{\ifmmode v_{\rm max} \else $v_{\rm max}$\fi} 

\newcommand{ \mumg  }{\ifmmode \mu\left(\mgii\right) \else $\mu\left(\mgii\right)$\fi}
\newcommand{ \fmg   }{\ifmmode f\left(\mgii\right) \else $f\left(\mgii\right)$\fi}
\newcommand{ \muciv }{\ifmmode \mu\left(\civ\right) \else $\mu\left(\civ\right)$\fi}
\newcommand{ \fciv  }{\ifmmode f\left(\civ\right) \else $f\left(\civ\right)$\fi}
\newcommand{  \auvo     }{\ifmmode \alpha_{\nu,{\rm UVO}} \else $\alpha_{\nu,{\rm UVO}}$\fi}
\newcommand{  \Ledd     }{\ifmmode L_{\rm Edd} \else $L_{\rm Edd}$\fi}
\newcommand{  \lamLlam  }{\ifmmode \lambda L_{\lambda} \else $\lambda L_{\lambda}$\fi}
\newcommand{  \lLl      }{\ifmmode \lambda L_{\lambda} \else $\lambda L_{\lambda}$\fi}
\newcommand{  \nuLnu    }{\ifmmode \nu L_{\nu} \else $\nu L_{\nu}$\fi}
\newcommand{  \nLn      }{\ifmmode \nu L_{\nu} \else $\nu L_{\nu}$\fi}
\newcommand{  \Luv      }{\ifmmode L_{1450} \else $L_{1450}$\fi}
\newcommand{  \lLop     }{\ifmmode \log\left(\Lop/\ergs\right) \else $\log\left(\Lop/\ergs\right)$\fi}
\newcommand{  \Lthree   }{\ifmmode L_{3000} \else $L_{3000}$\fi}
\newcommand{  \lLthree  }{\ifmmode \log\left(\Lthree/\ergs\right) \else $\log\left(\Lthree/\ergs\right)$\fi}
\newcommand{  \Lsix      }{\ifmmode L_{6200} \else $L_{6200}$\fi}
\newcommand{  \lLisx     }{\ifmmode \log\left(\Lop/\ergs\right) \else $\log\left(\Lop/\ergs\right)$\fi}
\newcommand{  \Lxray    }{\ifmmode L_{\rm X} \else $L_{\rm X}$\fi}

\newcommand{  \Lhard    }{\ifmmode L_{\rm 2-10} \else $L_{\rm 2-10}$\fi}
\newcommand{  \Lsoft    }{\ifmmode L_{\rm 0.5-2} \else $L_{\rm 0.5-2}$\fi}

\newcommand{\Fthree}{\ifmmode F_{3000} \else $F_{3000}$\fi}
\newcommand{\fuv}{\ifmmode f_{\lambda}\left(1450{\rm \AA}\right) \else $f_{\lambda}\left(1450 {\rm \AA}\right)$\fi}
\newcommand{\fthree}{\ifmmode f_{\lambda}\left(3000{\rm \AA}\right) \else $f_{\lambda}\left(3000{\rm \AA}\right)$\fi}
\newcommand{\fH}{\ifmmode f_{\lambda}\left(1.65\micron\right) \else
$f_{\lambda}\left(1.65\micron\right)$\fi}

\newcommand{\fbol}{\ifmmode f_{\rm bol} \else $f_{\rm bol}$\fi}
\newcommand{\fbolwv}{\ifmmode f_{\rm bol}\left(\lambda\right) \else $f_{\rm bol}\left(\lambda\right)$\fi}
\newcommand{\fbolopt}{\ifmmode f_{\rm bol}\left(5100{\rm \AA}\right) \else $f_{\rm bol}\left(5100{\rm \AA}\right)$\fi}
\newcommand{\fbolthree}{\ifmmode f_{\rm bol}\left(3000{\rm \AA}\right) \else $f_{\rm bol}\left(3000{\rm \AA}\right)$\fi}
\newcommand{\fboluv}{\ifmmode f_{\rm bol}\left(1450{\rm \AA}\right) \else $f_{\rm bol}\left(1450{\rm \AA}\right)$\fi}

\newcommand{\fbolbat}{\ifmmode f_{\rm bol}\left(14-150\,\kev\right) \else $f_{\rm bol}\left(14-150\,\kev\right)$\fi}
\newcommand{\fbolhard}{\ifmmode f_{\rm bol}\left(2-10\,\kev\right) \else $f_{\rm bol}\left(2-10\,\kev\right)$\fi}

\newcommand{\fobs}{\ifmmode f_{\rm obs} \else $f_{\rm obs}$\fi}

\newcommand{  \mbh      }{\ifmmode M_{\rm BH} \else $M_{\rm BH}$\fi}
\newcommand{  \lmbh     }{\ifmmode \log\left(\mbh/\Msun\right) \else $\log\left(\mbh/\Msun\right)$\fi} 
\newcommand{  \lledd    }{\ifmmode L/L_{\rm Edd} \else $L/L_{\rm Edd}$\fi}
\newcommand{  \mmedd    }{\ifmmode \dot{m}/\dot{m}_{\rm \,Edd} \else $\dot{m}/\dot{m}_{\rm \,Edd}$\fi}
\newcommand{  \Lbol     }{\ifmmode L_{\rm bol} \else $L_{\rm bol}$\fi}
\newcommand{  \lbol     }{\ifmmode L_{\rm bol} \else $L_{\rm bol}$\fi}
\newcommand{  \lLbol    }{\ifmmode \log\left(\Lbol/\ergs\right) \else $\log\left(\Lbol/\ergs\right)$\fi} 
\newcommand{  \Lagn     }{\ifmmode L_{\rm AGN} \else $L_{\rm AGN}$\fi}
\newcommand{  \lagn     }{\ifmmode L_{\rm AGN} \else $L_{\rm AGN}$\fi}

\newcommand{  \tgrow     }{\ifmmode t_{\rm growth} \else $t_{\rm growth}$\fi}
\newcommand{  \tAD     }{\ifmmode t_{\rm acc} \else $t_{\rm acc}$\fi}
\newcommand{  \tacc    }{\ifmmode t_{\rm acc} \else $t_{\rm acc}$\fi}
\newcommand{  \tUni      }{\ifmmode t_{\rm Universe} \else $t_{\rm Universe}$\fi}

\newcommand{  \Mdotin	}{\ifmmode \dot{M}_{\rm infall} \else $\dot{M}_{\rm infall}$\fi}
\newcommand{  \Mdotbh	}{\ifmmode \dot{M}_{\rm BH} \else $\dot{M}_{\rm BH}$\fi}
\newcommand{  \Mdotad	}{\ifmmode \dot{M}_{\rm AD} \else $\dot{M}_{\rm AD}$\fi}
\newcommand{  \Mdotacc	}{\ifmmode \dot{M}_{\rm acc} \else $\dot{M}_{\rm acc}$\fi}
\newcommand{  \Mdotthin	}{\ifmmode \dot{M}_{\rm thin} \else $\dot{M}_{\rm thin}$\fi}
\newcommand{  \Mdotdisk	}{\ifmmode \dot{M}_{\rm disk} \else $\dot{M}_{\rm disk}$\fi}

\newcommand{  \Mindot	}{\ifmmode \dot{M}_{\rm infall} \else $\dot{M}_{\rm infall}$\fi}
\newcommand{  \Mbhdot	}{\ifmmode \dot{M}_{\rm BH} \else $\dot{M}_{\rm BH}$\fi}
\newcommand{  \Maddot	}{\ifmmode \dot{M}_{\rm AD} \else $\dot{M}_{\rm AD}$\fi}
\newcommand{  \Maccdot	}{\ifmmode \dot{M}_{\rm acc} \else $\dot{M}_{\rm acc}$\fi}
\newcommand{  \Mthdot	}{\ifmmode \dot{M}_{\rm thin} \else $\dot{M}_{\rm thin}$\fi}
\newcommand{  \Mdsdot	}{\ifmmode \dot{M}_{\rm disk} \else $\dot{M}_{\rm disk}$\fi}

\newcommand{  \as	}{\ifmmode a_{\rm *} \else $a_{\rm *}$\fi}
\newcommand{  \avec	}{\ifmmode \vec{a}_{\rm *} \else $\vec{a}_{\rm *}$\fi}
\newcommand{  \re	}{\ifmmode \eta      	 \else $\eta$\fi}
\newcommand{  \RISCO	}{\ifmmode R_{\rm ISCO}  \else $R_{\rm ISCO}$\fi}

\newcommand{  \mseed    }{\ifmmode M_{\rm seed} \else $M_{\rm seed}$\fi}
\newcommand{  \mbul     }{\ifmmode M_{\rm bulge} \else $M_{\rm bulge}$\fi} 
\newcommand{  \mstar    }{\ifmmode M_{*} \else $M_{*}$\fi} 
\newcommand{  \mgal     }{\ifmmode M_{*} \else $M_{*}$\fi} 
\newcommand{  \mhost    }{\ifmmode M_{\rm host} \else $M_{\rm host}$\fi}
\newcommand{  \mmsmall  }{\ifmmode M_{\rm BH}/M_{*} \else $M_{\rm BH}/M_{*}$\fi}
\newcommand{  \mmlarge  }{\ifmmode M_{*}/M_{\rm BH} \else $M_{*}/M_{\rm BH}$\fi}

\newcommand{  \mmdotlarge}{\ifmmode \dot{M}_*/\Mbhdot \else $\dot{M}_*/\Mbhdot$\fi}
\newcommand{  \mmdotsmall}{\ifmmode \Mbhdot/\dot{M}_* \else $\Mbhdot/\dot{M}_*$\fi}

\newcommand{  \mmwp     }{\ifmmode \left(M_{*}/M_{\rm BH}\right) \else $\left(M_{*}/M_{\rm BH}\right)$\fi}
\newcommand{  \ml       }{\ifmmode M_{*}/L_{*} \else $M_{*}/L_{*}$\fi}
\newcommand{  \mlwp     }{\ifmmode \left(M_{*}/L\right) \else $\left(M_{*}/L\right)$\fi}
\newcommand{  \mlk      }{\ifmmode \left(M_{*}/L_{K}\right) \else $\left(M_{*}/L_{K}\right)$\fi}
\newcommand{  \sigs     }{\ifmmode \sigma_{*} \else $\sigma_{*}$\fi}
\newcommand{  \Reff     }{\ifmmode R_{\rm e} \else $R_{\rm e}$\fi}
\newcommand{  \Rvir     }{\ifmmode R_{\rm vir} \else $R_{\rm vir}$\fi}
\newcommand{  \Rtwo     }{\ifmmode R_{200} \else $R_{200}$\fi}
\newcommand{  \Rfive    }{\ifmmode R_{500} \else $R_{500}$\fi}
\newcommand{  \Rgrp     }{\ifmmode R_{\rm grp} \else $R_{\rm grp}$\fi}
\newcommand{  \nser     }{\ifmmode n_{\rm s} \else $n_{\rm s}$\fi}
\newcommand{  \LSF      }{\ifmmode L_{\rm SF}  \else $L_{\rm SF}$\fi}
\newcommand{  \LFIR     }{\ifmmode L_{\rm FIR} \else $L_{\rm FIR}$\fi}
\newcommand{  \Lfir     }{\ifmmode L_{\rm FIR} \else $L_{\rm FIR}$\fi}
\newcommand{  \LTIR     }{\ifmmode L_{\rm TIR} \else $L_{\rm TIR}$\fi}
\newcommand{  \Ltir     }{\ifmmode L_{\rm TIR} \else $L_{\rm TIR}$\fi}

\newcommand{  \mdyn     }{\ifmmode M_{\rm dyn} \else $M_{\rm dyn}$\fi} 
\newcommand{  \mgas     }{\ifmmode M_{\rm gas} \else $M_{\rm gas}$\fi} 
\newcommand{  \mh       }{\ifmmode M_{\rm h} \else $M_{\rm h}$\fi}
\newcommand{  \mhalo    }{\ifmmode M_{\rm halo} \else $M_{\rm halo}$\fi}
\newcommand{  \sfr      }{\ifmmode {\rm SFR} \else SFR\fi}

\newcommand{ \Lcii     }{\ifmmode L_{\cii} \else $L_{\cii}$\fi}
\newcommand{ \fwcii  }{\ifmmode {\rm FWHM}\cii \else FWHM\cii\fi}

\newcommand  {\RBLR}        {\hbox{$ {R_{\rm BLR}} $}}

\newcommand{\bj}{\ifmmode b_{\rm J} \else $b_{\rm J}$\fi}

\newcommand{\iab}{\ifmmode i_{\rm AB} \else $i_{\rm AB}$\fi}

\newcommand{\jab}{\ifmmode J_{\rm AB} \else $J_{\rm AB}$\fi}
\newcommand{\hab}{\ifmmode H_{\rm AB} \else $H_{\rm AB}$\fi}
\newcommand{\kab}{\ifmmode K_{\rm AB} \else $K_{\rm AB}$\fi}

\newcommand{\jveg}{\ifmmode J_{\rm Vega} \else $J_{\rm Vega}$\fi}
\newcommand{\hveg}{\ifmmode H_{\rm Vega} \else $H_{\rm Vega}$\fi}
\newcommand{\kveg}{\ifmmode K_{\rm Vega} \else $K_{\rm Vega}$\fi}

\newcommand{  \Chisq    }{\ifmmode \chi^{2} \else $\chi^{2}$}
\newcommand{  \nelec    }{\ifmmode n_{e} \else $n_{e}$\fi}     % electron density
\newcommand{  \nh       }{\ifmmode n_{\rm H} \else $n_{\rm H}$\fi}     % hydrogen density
\newcommand{  \Ncol     }{\ifmmode N_{\rm col} \else $N_{\rm col}$\fi} % column density
\newcommand{  \NH       }{\ifmmode N_{\rm H} \else $N_{\rm H}$\fi}     % column density

\def\sun{\hbox{$\odot$}}

\def\arcsec{\hbox{$^{\prime\prime}$}}

\DeclareRobustCommand{\ion}[2]{%
\relax\ifmmode
\ifx\testbx\f@series
{\mathbf{#1\,\mathsc{#2}}}\else
{\mathrm{#1\,\mathsc{#2}}}\fi
\else\textup{#1\,{\mdseries\textsc{#2}}}%
\fi}

\usepackage[latin1]{inputenc}
\usepackage[T1]{fontenc}
\usepackage{soul}

\received{August 15, 2022}
\revised{October 13, 2022}
\accepted{October 14, 2022}
\submitjournal{ApJL}

\shorttitle{The CLAGN \target}
\shortauthors{Zeltyn et al.}
\graphicspath{{./}{figures/}}
\begin{document}

\title{A Transient ``Changing-look'' Active Galactic Nucleus Resolved on Month Timescales from First-year Sloan Digital Sky Survey V Data}

\author[0000-0002-7817-0099]{Grisha Zeltyn}
\affiliation{School of Physics and Astronomy, Tel Aviv University, Tel Aviv 69978, Israel}

\author[0000-0002-3683-7297]{Benny Trakhtenbrot}
\affiliation{School of Physics and Astronomy, Tel Aviv University, Tel Aviv 69978, Israel}

\author[0000-0002-3719-940X]{Michael Eracleous}
\affiliation{Department of Astronomy \& Astrophysics and Institute for Gravitation and the Cosmos, The Pennsylvania State University, 525 Davey Lab, University Park, PA 16802, USA}

\author[0000-0001-8557-2822]{Jessie Runnoe}
\affiliation{Department of Physics and Astronomy, Vanderbilt University, Nashville, TN 37235, USA}

\author[0000-0002-1410-0470]{Jonathan R. Trump}
\affiliation{Department of Physics, 196 Auditorium Road, Unit 3046, University of Connecticut, Storrs, CT 06269, USA}

\author[0000-0002-7541-9565]{Jonathan Stern}
\affiliation{School of Physics and Astronomy, Tel Aviv University, Tel Aviv 69978, Israel}

\author[0000-0003-1659-7035]{Yue Shen}
\affiliation{Department of Astronomy, University of Illinois at Urbana-Champaign, Urbana, IL 61801, USA}
\affiliation{National Center for Supercomputing Applications, University of Illinois at Urbana-Champaign, Urbana, IL 61801, USA}

\author[0000-0002-8606-6961]{Lorena Hern\'andez-Garc\'ia}
\affiliation{Millennium Institute of Astrophysics (MAS), Nuncio Monse\~{n}or S\'otero Sanz 100, Providencia, Santiago, Chile}
\affiliation{Instituto de F\'isica y Astronom\'ia, Facultad de Ciencias, Universidad de Valpara\'iso, Gran Breta\~{n}a 1111, Valpara\'iso, Chile}

\author[0000-0002-8686-8737]{Franz E.\ Bauer}
\affiliation{Instituto de Astrof\'isica and Centro de Astroingenier\'ia, Facultad de F\'isica, Pontificia Universidad Cat\'olica de Chile, Casilla 306, Santiago 22, Chile}
\affiliation{Millennium Institute of Astrophysics (MAS), Nuncio Monse\~{n}or S\'otero Sanz 100, Providencia, Santiago, Chile}
\affiliation{Space Science Institute, 4750 Walnut Street, Suite 205, Boulder, CO 80301, USA}

\author[0000-0002-6893-3742]{Qian Yang}
\affiliation{Department of Astronomy, University of Illinois at Urbana-Champaign, Urbana, IL 61801, USA}
\affiliation{Center for Astrophysics, Harvard \& Smithsonian, 60 Garden Street, Cambridge, MA 02138, USA}

\author[0000-0002-4459-9233]{Tom Dwelly}
\affiliation{Max-Planck-Institut f{\"u}r extraterrestrische Physik, Giessenbachstra\ss{}e, D-85748 Garching, Germany}

\author[0000-0001-5231-2645]{Claudio Ricci}
\affiliation{N\'ucleo de Astronom\'ia de la Facultad de Ingenier\'ia, Universidad Diego Portales, Av. Ej\'ercito Libertador 441, Santiago, Chile}
\affiliation{Kavli Institute for Astronomy and Astrophysics, Peking University, Beijing 100871, People's Republic of China}

\author[0000-0002-8179-9445]{Paul Green}
\affiliation{Center for Astrophysics, Harvard \& Smithsonian, 60 Garden Street, Cambridge, MA 02138, USA}

\author[0000-0002-6404-9562]{Scott F. Anderson}
\affiliation{Astronomy Department, University of Washington, Box 351580, Seattle, WA 98195, USA}

\author[0000-0002-9508-3667]{Roberto J. Assef}
\affiliation{N\'ucleo de Astronom\'ia de la Facultad de Ingenier\'ia, Universidad Diego Portales, Av. Ej\'ercito Libertador 441, Santiago, Chile}

\author[0000-0002-5063-0751]{Muryel Guolo}
\affiliation{Department of Physics and Astronomy, Johns Hopkins University, 3400 North Charles Street, Baltimore MD 21218, USA}

\author[0000-0003-3422-2202]{Chelsea MacLeod}
\affiliation{BlackSky, 1505 Westlake Avenue North \#600, Seattle, WA 98109, USA}

%%%%%%%%%%%%%%%%%%%%%%%%%%%%%%%%%%%%%%%%%%%%%

\author[0000-0001-9776-9227]{Megan C. Davis}
\affiliation{Department of Physics, 196 Auditorium Road, Unit 3046, University of Connecticut, Storrs, CT 06269, USA}

\author[0000-0001-8032-2971]{Logan Fries}
\affiliation{Department of Physics, 196 Auditorium Road, Unit 3046, University of Connecticut, Storrs, CT 06269, USA}

\author[0000-0003-3703-5154]{Suvi Gezari}
\affiliation{Space Telescope Science Institute, 3700 San Martin Drive, Baltimore, MD 21218, USA}

\author[0000-0001-9440-8872]{Norman A. Grogin}
\affiliation{Space Telescope Science Institute, 3700 San Martin Drive, Baltimore, MD 21218, USA}

\author{David Homan}
\affiliation{Leibniz-Institut f\"{u}r Astrophysik Potsdam, An der Sternwarte 16, D-14482 Potsdam, Germany}

\author[0000-0002-6610-2048]{Anton M. Koekemoer}
\affiliation{Space Telescope Science Institute, 3700 San Martin Drive, Baltimore, MD 21218, USA}

\author{Mirko Krumpe}
\affiliation{Leibniz-Institut f\"{u}r Astrophysik Potsdam, An der Sternwarte 16, D-14482 Potsdam, Germany}

\author[0000-0002-5907-3330]{Stephanie LaMassa}
\affiliation{Space Telescope Science Institute, 3700 San Martin Drive, Baltimore, MD 21218, USA}

\author[0000-0003-0049-5210]{Xin Liu}
\affiliation{Department of Astronomy, University of Illinois at Urbana-Champaign, Urbana, IL 61801, USA}
\affiliation{National Center for Supercomputing Applications, University of Illinois at Urbana-Champaign, Urbana, IL 61801, USA}

\author[0000-0002-0761-0130]{Andrea Merloni}
\affiliation{Max-Planck-Institut f{\"u}r extraterrestrische Physik, Giessenbachstra\ss{}e, D-85748 Garching, Germany}

\author[0000-0002-7843-7689]{Mary Loli Mart\'inez-Aldama}
\affiliation{Instituto de F\'isica y Astronom\'ia, Facultad de Ciencias, Universidad de Valpara\'iso, Gran Breta\~{n}a 1111, Valpara\'iso, Chile}
\affiliation{Departamento de Astronom\'ia, Universidad de Chile, Casilla 36D, Santiago, Chile}

\author[0000-0001-7240-7449]{Donald P.\ Schneider} 
\affiliation{Department of Astronomy \& Astrophysics and Institute for Gravitation and the Cosmos, The Pennsylvania State University, 525 Davey Lab, University Park, PA 16802, USA}

\author[0000-0001-8433-550X]{Matthew J. Temple}
\affil{N\'ucleo de Astronom\'ia de la Facultad de Ingenier\'ia, Universidad Diego Portales, Av. Ej\'ercito Libertador 441, Santiago, Chile}

%%%%%%%%%%%%%%%%%%%%%%%%%%%%%%%%%%%%%%%%%%%%%

\author[0000-0002-8725-1069]{Joel R. Brownstein}
\affiliation{Department of Physics and Astronomy, University of Utah, 115 S. 1400 East, Salt Lake City, UT 84112, USA}

\author[0000-0002-9790-6313]{Hector Ibarra-Medel}
\affiliation{Department of Astronomy, University of Illinois at Urbana-Champaign, Urbana, IL 61801, USA}
\affiliation{Instituto de Astronom\'ia y Ciencias Planetarias, Universidad de Atacama, Copayapu 485, Copiap\'o, Chile}

%%%%%%%%%%%%%%%%%%%%%%%%%%%%%%%%%%%%%%%%%%%%%

\author[0000-0003-0035-6659]{Jamison Burke}
\author[0000-0002-7472-1279]{Craig Pellegrino}
\affil{Las Cumbres Observatory, 6740 Cortona Drive, Suite 102, Goleta,
CA 93117-5575, USA}
\affil{Department of Physics, University of California, Santa Barbara, CA
93106-9530, USA}

%%%%%%%%%%%%%%%%%%%%%%%%%%%%%%%%%%%%%%%%%%%%%
\author[0000-0001-9852-1610]{Juna A. Kollmeier}
\affiliation{The Observatories of the Carnegie Institution for Science, 813 Santa Barbara Street, Pasadena, CA 91101, USA}
\affiliation{Canadian Institute for Theoretical Astrophysics, 60 Saint George Street, Toronto, ON M5S 3H8, Canada}

%%%%%%%%%%%%%%%%%%%%%%%%%%%%%%%%%%%%%%%%%%%%%

\correspondingauthor{Grisha Zeltyn}
\email{grishazeltyn@tauex.tau.ac.il, benny@astro.tau.ac.il}

\begin{abstract}
We report the discovery of a new ``changing-look'' active galactic nucleus (CLAGN) event, in the quasar SDSS J162829.17+432948.5 at $z=0.2603$, 
identified through repeat spectroscopy from the fifth Sloan Digital Sky Survey (SDSS-V).
Optical photometry taken during 2020--2021 shows a dramatic dimming of $\Delta g{\approx}1 \rm{mag}$, followed by a rapid recovery on a timescale of several months, with the $\lesssim$2 month period of rebrightening captured in new SDSS-V and Las Cumbres Observatory spectroscopy.
This is one of the fastest CLAGN transitions observed to date.
Archival observations suggest that the object experienced a much more gradual dimming over the period of 2011--2013.
Our spectroscopy shows that the photometric changes were accompanied by dramatic variations in the quasar-like continuum and broad-line emission.
The excellent agreement between the pre- and postdip photometric and spectroscopic appearances of the source, 
as well as the fact that the dimmest spectra can be reproduced by applying a single extinction law to the brighter spectral states, 
favor a variable line-of-sight obscuration as the driver of the observed transitions. 
Such an interpretation faces several theoretical challenges, and thus an alternative accretion-driven scenario cannot be excluded.
The recent events observed in this quasar highlight the importance of spectroscopic monitoring of large active galactic nucleus samples on weeks-to-months timescales, which the SDSS-V is designed to achieve.
\end{abstract}

\keywords{Supermassive black holes (1663), Quasars (1319), Active galactic nuclei (16), Transient sources (1861)}

%%%%%%%%%%%%%%%%%%%%%%%%%%%%%%%%%%%%%%%%%%%%%%%%%%%%%%%%%%%%%%%%
\section{Introduction}
\label{sec:intro}

In recent years, optical time-domain photometric and spectroscopic wide-area surveys have revealed a multitude of highly variable active galactic nuclei (AGNs). 
Among those, ``changing-look'' AGNs (CLAGNs hereafter) identified in the rest-frame UV-optical regime are events that show significant changes of the blue continuum and/or broad emission lines (BELs) that are typical of unobscured AGNs, often resulting in a transition between AGN-dominated and host-dominated spectral appearances.
\footnote{ Throughout this Letter, we use the term CLAGN to refer to {\it all} AGNs that show such spectral transitions, regardless of the physical mechanisms driving these changes.} 
Over the past decade, many such transitions have been discovered among highly luminous AGNs (i.e., quasars), starting from a few prototypical cases \cite[e.g.,][]{LaMassa15,Runnoe16}, followed by more sizable samples \citep[e.g.,][]{Yang18, MacLeod19, Green22}.
The nature of the surveys used to identify these events means that the extreme variability is typically traced over timescales of years. 
To date, only a few examples clearly show shorter timescale {\it spectral} transitions in the UV-optical regime \cite[]{Guo16, Trakhtenbrot19, Ross20}. 
In several more cases, photometric monitoring programs imply that spectral transitions have in fact occurred over timescales of $<$1 yr \cite[e.g.,][]{Gezari17, Yang18, Frederick19, Yan19, Green22}.

Identifying the physical mechanisms that drive the extreme variability observed in CLAGNs can provide information about the structure and physics of the accretion flow, broad-line region (BLR) and other key AGN components, and yield insights into intermittent supermassive black hole (SMBH) growth. 
The mechanisms driving UV-optical CLAGN transitions, however, remain unclear, with most studies favoring explanations related to changes in the (ionizing) continuum radiation power due to variations in the accretion flow \cite[e.g.,][]{LaMassa15, Ruan16, Runnoe16, Sheng17, Rumbaugh18, Stern18, Trakhtenbrot19, Guolo21}. 
Alternatively, some of the spectral transitions could result from changes in obscuring dust/gas along our line of sight (\los). Many studies of extremely variable AGNs identified in the X-rays have shown decisive evidence for variable obscuration of the central engines of some AGNs
\citep[e.g.,][and references therein]{Risaliti05, Maiolino10,Markowitz14,Hernandez17, Liu22}. 
In addition, variations to the \los\ obscuration of the continuum source and the BLR have been inferred from the variable optical spectra of a few Seyfert 1.8 and 1.9 galaxies (e.g., \citealt{Goodrich95}; see also \citealt{GH18} and references therein).

In this Letter we report a newly discovered CLAGN, \fulltarget\ (hereafter \target), which displays recurring dramatic spectral changes over relatively short timescales, as well as transitions between seemingly distinct spectral states. 
After describing the observations and spectral decomposition methods (Section~\ref{sec:obs_methods}), we discuss several possible physical interpretations of the observed changes and conclude that the data may be better explained by variable obscuration (Section~\ref{sec:dicussion}).
These observations may offer the best evidence to date for variable obscuration driving a CLAGN transition in a luminous quasar, and, more generally, one of the fastest CLAGN transitions traced in both photometry and spectroscopy. 
A summary of our results is presented in Section~\ref{sec:summary}. 

Throughout this work, we adopt a flat $\Lambda$ cold dark matter cosmology with $H_0=70\,\kms\,\rm{Mpc}^{-1}$ and $\Omega_{\rm{m}}=0.3$.

%%%%%%%%%%%%%%%%%%%%%%%%%%%%%%%%%%%%%%%%%%%%%%%%%%%%%%%%%%%%%%%%
\section{Observations and Analysis}
\label{sec:obs_methods}

\subsection{Multiepoch Optical Spectroscopy}
\label{subsec:obs_spec}

\begin{figure*}
    \centering
    \includegraphics[width=1.0\textwidth]{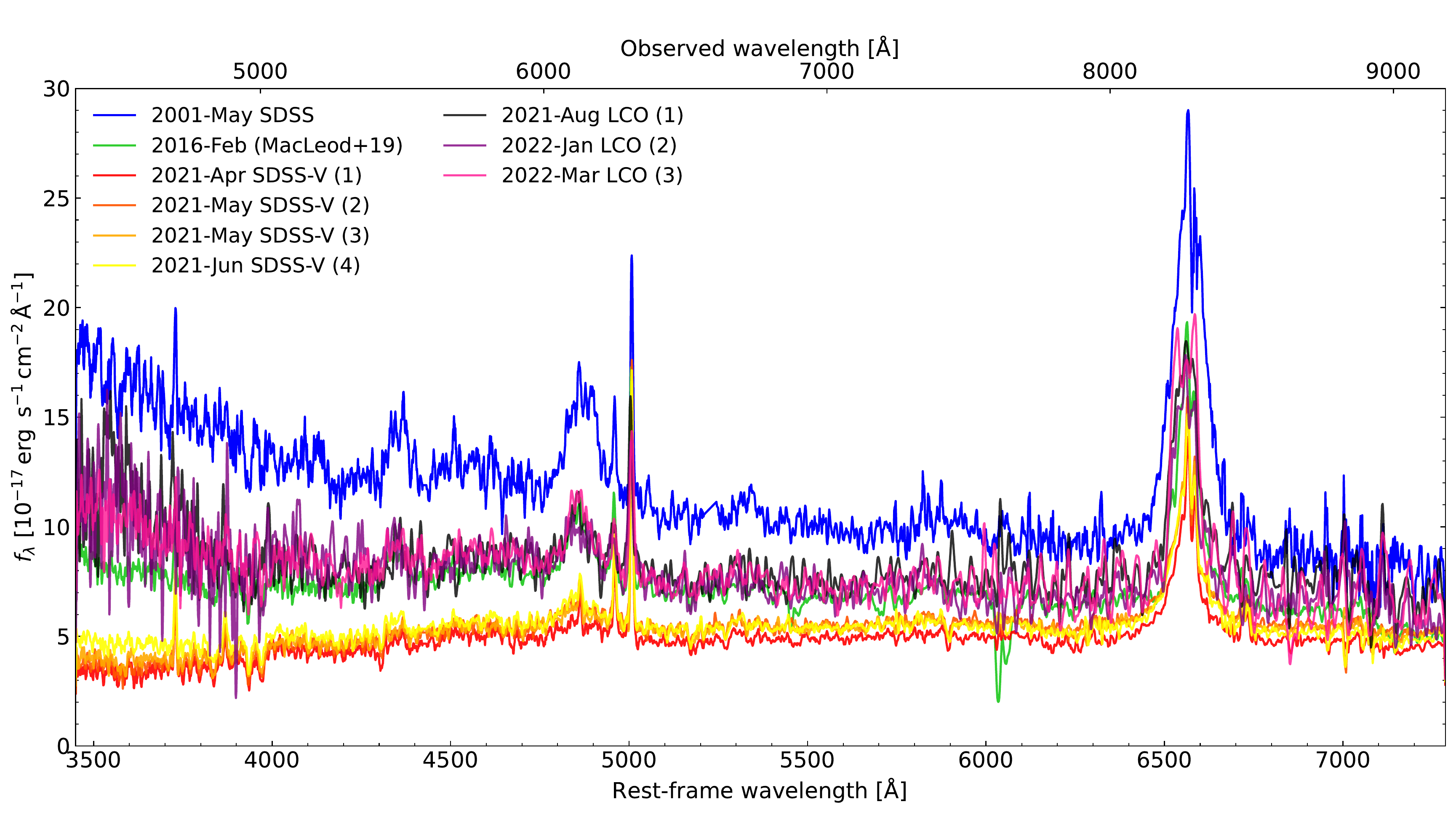}
    \caption{Multiepoch optical spectra of \target. 
    This AGN shows a gradual dimming of its quasar-like continuum and Balmer BELs across $\sim20$ yr, through the 2001 bright-state Sloan Digital Sky Survey (SDSS) spectrum, the 2016 intermediate-state WHT spectrum \citep[]{MacLeod19}, and the 2021 dim-state SDSS-V spectra. Subsequent observations performed by Las Cumbres Observatory (LCO) revealed a brightening to an intermediate state in $\lesssim2$ months.}
    \label{fig:spectra}
\end{figure*}

Our key observational dataset is multiepoch medium-resolution optical spectroscopy of the AGN \target, at $z=0.2603$ (at a luminosity distance of $\approx1320\,\rm{Mpc}$ and an angular scale of $\approx$4 kpc/\arcsec), collected over a span of over two decades.
% , which we present below in chronological order.
The spectra used for our main analysis are shown in Fig.~\ref{fig:spectra}; details about the observations are provided in Table~\ref{tab:fit_params}.

The first spectrum was obtained in 2001 May as part of the legacy Sloan Digital Sky Survey \cite[SDSS;][]{York2000}. %Ahumada20
This spectrum shows a blue continuum and broad Balmer emission lines typical of unobscured AGNs, with a quasar-like continuum luminosity ($\log(\lambda L_{\lambda}[5100\,{\rm \AA}]/\ergs)\simeq44$; see Section~\ref{subsec:spec_analysis})
In what follows, we refer to this 2001 spectrum as the ``bright state.''

The second spectrum was obtained following the identification of \target\ as a highly variable source based on SDSS and Pan-STARRS1 \cite[PS1;][]{Chambers16} photometry, as part of the study by \citet[][see their Table~2]{MacLeod19}. The spectrum was acquired in 2016 February using the Intermediate dispersion Spectrograph and Imaging System (ISIS) mounted on the 4.2 m William Herschel Telescope (WHT) at the Roque de los Muchachos Observatory. 
While the 2016 spectrum showed a fainter continuum and broad Balmer-line emission compared to the 2001 bright-state spectrum, the analysis of \citet[]{MacLeod19} found insignificant line variability.
We refer to this spectrum as the ``intermediate state.''

During 2021 April--June, four more spectra were obtained as part of the Black Hole Mapper program (S.F. Anderson et al., in preparation) within the ongoing fifth generation of the SDSS (SDSS-V; \citealt{Kollmeier17}, J.A. Kollmeier et al., in preparation) obtained with the plate-based fiber-fed BOSS spectrograph \citep{Smee13} mounted on the SDSS 2.5 m Telescope \citep{Gunn06} at the Apache Point Observatory. 
The four visits are part of a dedicated SDSS-V subprogram that aims to monitor previously known quasars over timescales of weeks to years.
The four SDSS-V spectra, taken within a period of $\approx$2 months, are all consistent with each other, but differ dramatically from earlier spectra, suggesting the SDSS-V spectra have captured a CLAGN event.
Specifically, the quasar appeared much fainter than the previous (2001 and 2016) observations, showing an almost complete disappearance of the quasar-like blue continuum and broad \Hb line emission. 
We refer to the state captured by these spectra as the ``dim state.''

Following the identification of the spectral variability of \target, \NLCO\ further spectra were acquired with the FLOYDS spectrograph mounted on the 2 m Faulkes Telescope North at Haleakala, Hawaii, which is a part of the Las Cumbres Observatory network \cite[LCO;][]{Brown13}. The spectra were obtained ${\approx}2$, 8, and 10 months after the latest dim-state SDSS-V spectrum. Surprisingly, these spectra revealed a reappearance of the blue continuum and the prominent broad \Hb line emission, reverting to a state that closely resembles the 2016 intermediate-state spectrum. 

Subsequent spectroscopy obtained with the Astrophysical Research Consortium 3.5 m Telescope (Apache Point Observatory), the Hobby--Eberly Telescope (McDonald Observatory), and the continuing SDSS-V program all confirm that \target\ indeed retains its intermediate-state spectral appearance, as of 2022 May 21. 
The absolute flux calibration of these spectra, however, has large uncertainties, and these spectra are thus not included in our main analysis.

\subsection{Multiepoch Imaging}
\label{subsec:obs_img}

\begin{figure*}
    \centering
    \includegraphics[width=1.0\textwidth]{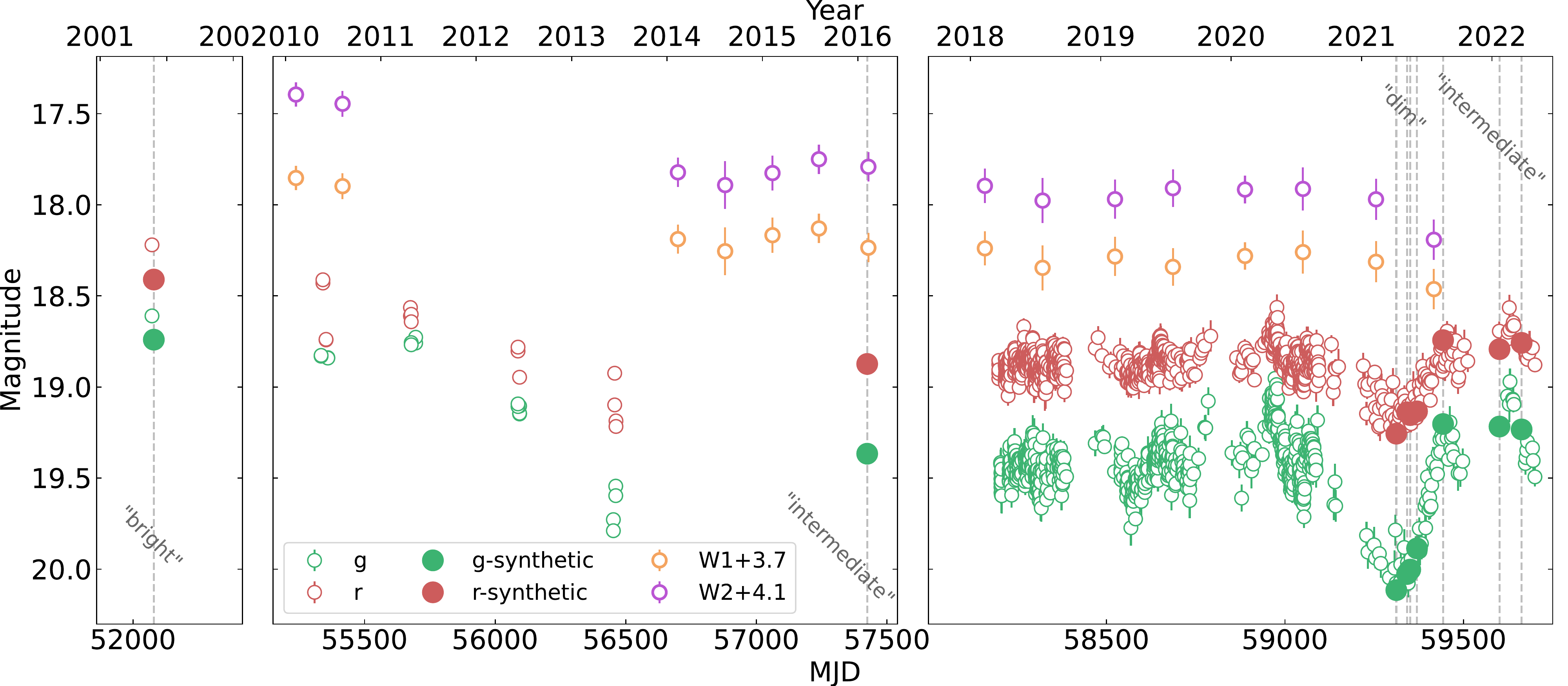}
    \caption{Optical light curves of \target\ obtained from SDSS (left), Pan-STARRS (middle), and ZTF (right) imaging, in the $g$ and $r$ bands (open circles; see legend) over a period of two decades. 
    For clarity, only measurements with magnitude uncertainties less than 0.1 mag are displayed.
    Dashed vertical lines mark the epochs of the spectra shown in Fig.~\ref{fig:spectra}, and the corresponding synthetic photometry is marked with filled circles.
    Note the fast, $<$2 month transitions observed within ZTF during 2021, and the slower dimming observed within Pan-STARRS, during 2012--2014.
    The middle and right panels also show light curves obtained from WISE, in the $W1$ and $W2$ bands, where each data point shows the median magnitude at the given epoch. Note the vertical offset of the WISE data (for clarity; see legend).
    }
    \label{fig:photometry}
\end{figure*}

Figure~\ref{fig:photometry} shows $g$- and $r$-band photometry of \target. The photometric measurements were obtained from publicly available legacy SDSS, PS1 \cite[]{Flewelling20}, and  Zwicky Transient Facility (ZTF; \citealt{Masci19}) data.\footnote{We applied quality metrics following ZTF guidelines (\url{https://irsa.ipac.caltech.edu/data/ZTF/docs/ztf_forced_photometry.pdf}) and using procedures developed by the ALeRCE team \cite[]{Forster21}, which removes bad weather data and applies a color correction.}
We also show synthetic photometry derived from each of the \target\ spectra, by convolving the spectra with the respective filter curves. 
The PS1 photometry reveals a steady dimming, of $\Delta g \approx 1$ mag, between 2011 and 2013, consistent with the trend observed between the 2001 and the 2016 spectra.
In addition, the ZTF (forced) photometry clearly confirms the rapid variations seen between the 2021 SDSS-V and LCO spectra, and clearly demonstrates that \target\ became fainter by another $\Delta g \approx 1$ mag between 2021 March and June, following a period of limited variability during 2018--2021. 
Such changes of $\Delta g \approx 1$ mag on a timescale of a few months are inconsistent with observations of typical AGNs, which require much longer timescales for similarly extreme variability to occur \cite[see, e.g.,][and references therein]{Rumbaugh18}.

In addition, Figure~\ref{fig:photometry} displays mid-infrared photometry of \target\ obtained from the Wide-field Infrared Survey Explorer (WISE; \citealt{Wright10}), in the $W1$ and $W2$ bands ($\approx$3.4 and 4.6 \mic, respectively). The WISE light curves reveal a dimming of $\approx0.4$ mag between 2011 and 2014 in both bands, and may also show hints of a slightly weaker dimming in the most recent epoch available at the time of writing, in 2021 July.

\subsection{Spectral Analysis}
\label{subsec:spec_analysis}

We employed the \texttt{PyQSOFit} code \citep[]{QSOFit,Shen19} to obtain key spectral measurements of \target, in each of the various epochs observed. 
We first verified that the observed fluxes of the narrow \OIII\ line vary by less than $\approx$15\% between the different spectra (see Table \ref{tab:fit_params}). 
Considering also the relatively small angular size of \target\ (compared with the apertures used; see Table~\ref{tab:fit_params}), and the consistency between the synthetic and imaging photometry (Fig.~\ref{fig:photometry}), we chose to analyze the spectra without any additional scaling. %
In what follows, we briefly describe the key steps of the \texttt{PyQSOFit} fitting process. 

All spectra were shifted to the rest frame and corrected for Galactic extinction using the \citet[]{SFD98} dust maps ($A_V\simeq0.02$) and a Milky Way (MW) extinction law \cite[see][and references therein]{SN20}. 
The faintest, most host-galaxy dominated, SDSS-V spectrum (taken in 2021 April) was then decomposed to extract the host-galaxy spectrum, relying on the template-based principal component analysis approach implemented within \texttt{PyQSOFit}.
We then subtracted this faint-state-based host spectrum from all spectra of \target.
We verified that our spectral decomposition reproduces the stellar absorption features observed in the dim- and intermediate-state spectra. 
For the bright state, however, matching the host-galaxy features is less robust, as expected given the quasar-dominated continuum. 
We stress that our main analysis focuses on the intermediate--dim--intermediate transition.
The continuum of the host-subtracted spectra was then modeled as a combination of a simple power law, a Balmer continuum model, 
and an optical \ion{Fe}{ii} composite model. We did not add a polynomial component to the continuum modeling, as it did not significantly affect our result.
We fitted all broad and narrow emission lines with one Gaussian (each), except for the broad \Ha line, which required two Gaussians to produce an acceptable fit. We verified that fixing the fit parameters of the narrow emission lines across all spectra did not significantly affect our results.

\texttt{PyQSOFit} was used to obtain the FWHM, EW, and flux of the broad \Ha and \Hb emission lines, as well as the AGN-only luminosity at $5100\,\AA$, $\lambda L_{\lambda}[5100\,{\rm \AA}]$ (\Lop\ hereafter), for each of the epochs. 
These quantities are listed in Table \ref{tab:fit_params}.
The uncertainties were obtained through a Monte Carlo refitting approach, using 100 realizations for each spectrum, relying on the corresponding error spectra.
The best-fit spectral parameters derived from the fit of the \Hb\ complex in the (host-subtracted) ``bright-state'' spectrum and the single-epoch prescription for black hole masses imply $\log(\mbh/\Msun)\simeq8.2$ \cite[see][and references therein]{Mejia-Restrepo22}.  Assuming an optical-to-bolometric correction of $\Lbol/\Lop=9.3$ \citep[]{Shen11} yields a mass accretion rate of $\dot{M}=\Lbol/(\eta c^2)\simeq0.16\,\Msunyr$ (assuming $\eta=0.1$) and an Eddington ratio of $\lledd\equiv\Lbol/(1.5\times10^{38} \times [\mbh/\Msun])\simeq0.03$. 
The (host-subtracted) faintest SDSS-V spectrum leads to $\dot{M}\simeq0.033\,\Msunyr$ and $\lledd\simeq0.007$, under the same assumptions.
All these quantities carry significant systematic uncertainties, of $\gtrsim$0.3 dex.

\begin{figure}
    \centering
    \includegraphics[width=1.0\columnwidth]{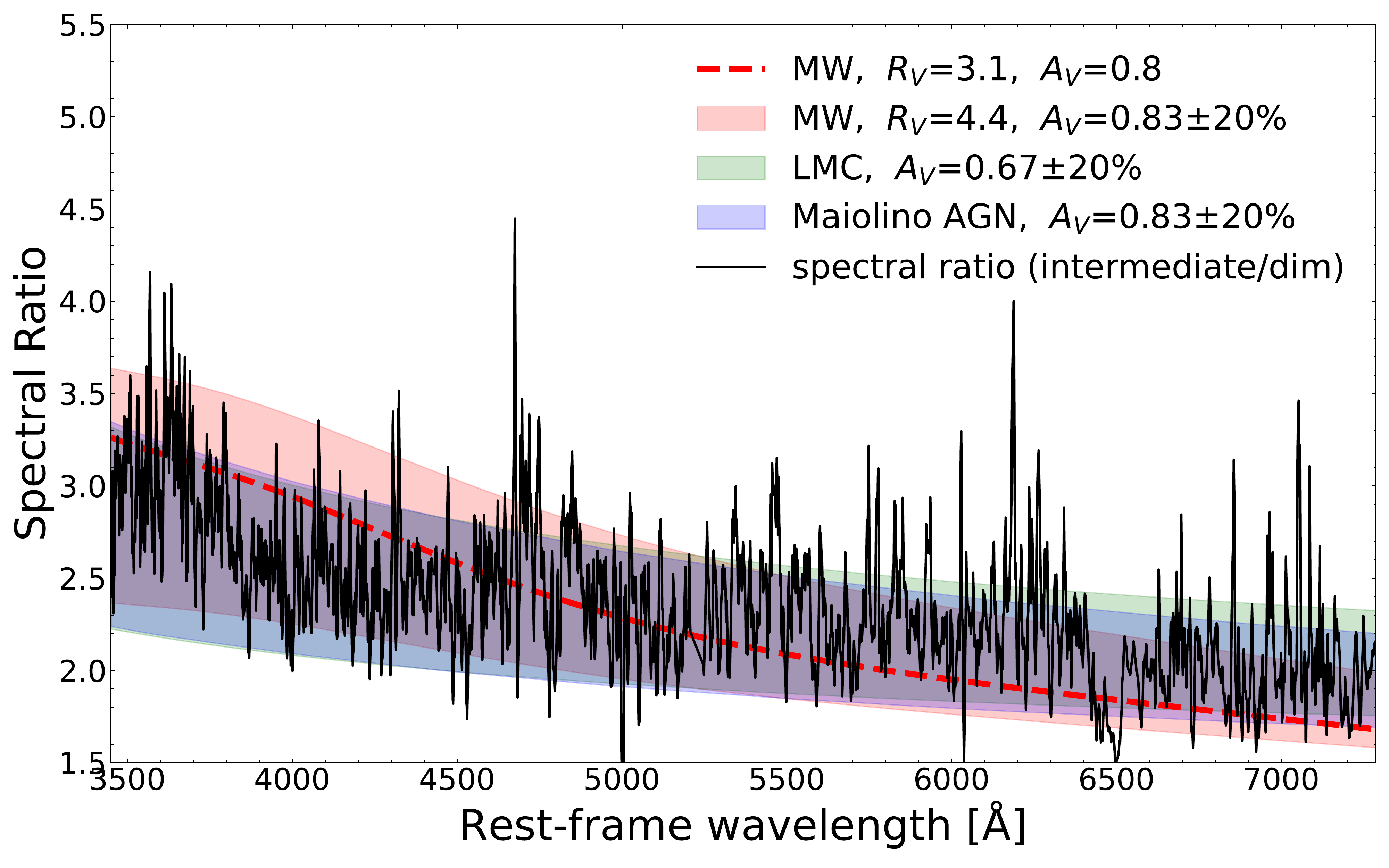}
    \caption{The spectral ratio between the 2016 intermediate state and 2021 dim state of J1628+4329 (black), smoothed over 7 pixels. The colored areas indicate different extinction laws: Milky Way (MW)-like extinction with $R_V=4.4$ and LMC-like extinction \citep[red and green, respectively; see][and references therein]{SN20}, and ``Maiolino-AGN'' extinction \citep[blue;][]{Maiolino01,Li07}. A range of possible $A_V$ values for each extinction law is shown. Steeper extinction laws, e.g., MW-like with $R_V=3.1$, are not consistent with the spectral ratio (see dashed line).}
    \label{fig:ratio}
\end{figure}

Figure~\ref{fig:ratio} presents the spectral ratio between the 2016 intermediate-state and the 2021 dim-state spectra, calculated for the host- and narrow-line-subtracted spectra (i.e., the quasar-like continuum and Balmer BELs). 
The spectral ratio demonstrates two key aspects of the changes seen in \target: 
(1) the redder nature of the dim state (i.e., the ratio spectrum rises toward shorter wavelengths); 
and (2) the overall smooth variation of the spectral ratio across the entire wavelength range, particularly in the spectral bands adjacent to the Balmer BELs (i.e., the ratio spectrum itself shows only weak and noisy features coincident with the Balmer-line wavelengths, and does not show features that resemble the broad Balmer emission profiles themselves). 
The implications of this figure are further explored in Section \ref{sec:dicussion}.

\begin{deluxetable*}{llllcccccccccc}
% \tablenum{1}
\label{tab:fit_params}
\tablecaption{Key Multiepoch Spectral Measurements for \target.}
\tablewidth{0pt}
\tabletypesize{\scriptsize}
\tablehead{
\colhead{Date} & \colhead{MJD} & \colhead{Telescope}  & \colhead{$d$\tablenotemark{a}} & \colhead{\fwha} & \colhead{\ewha} & \colhead{$F$(\Ha)/$10^{-17}$} & \colhead{\fwhb} & \colhead{\ewhb} & \colhead{$F$(\Hb)/$10^{-17}$} & \colhead{$\log\Lop$}
& \colhead{$F(\oiii)/10^{-17}$} & \colhead{State}
\\
\colhead{} & \colhead{} & \colhead{} & \colhead{(\arcsec)} & \colhead{(\kms)} & \colhead{(\AA)} & \colhead{($\ergs\rm{cm}^{-2}$)} & \colhead{(\kms)} & \colhead{(\AA)} &  \colhead{($\ergs\rm{cm}^{-2}$)} & \colhead{(\ergs)}
& \colhead{($\ergs\rm{cm}^{-2}$)} & \colhead{}
}
\startdata
2001 May 28 & 52057 & SDSS 2.5m/SDSS & 3 & 5011$\pm$93 & 425$\pm$28 & 2770$\pm$180 & 5790$\pm$190 & 63$\pm$2 & 622$\pm$17 & 44.0$\pm$0.005 & $126^{+5}_{-6}$ & ``Bright'' \\ \hline
2016 Feb 6 & 57424 & WHT 4.2m/ISIS & 1 &3750$\pm$120 & 337$\pm$12 & 1157$\pm$41 & 5450$\pm$160 & 67$\pm$1 & 347$\pm$7 & 43.7$\pm$0.004 & 130$\pm$3 & ``Intermediate'' \\ \hline
2021 Apr 8 & 59312 & SDSS 2.5m/BOSS & 2 &4130$\pm$140 & 423$\pm$11 & 681$\pm$17 & 6700$\pm$1200 & 46$\pm$5 & 93$\pm$10 & 43.3$\pm$0.006 & $117^{+3}_{-2}$ & ``Dim''\\
2021 May 8 & 59342 & SDSS 2.5m/BOSS & 2 &4330$\pm$150 & 298$\pm$7 & 744$\pm$16 & 8800$\pm$980 & 71$\pm$6 & 178$\pm$14 & 43.4$\pm$0.008 & 122$\pm$3\\
2021 May 17 & 59351 & SDSS 2.5m/BOSS & 2 & 4250$\pm$140 & 323$\pm$7 & 763$\pm$15 & 6400$\pm$1200 & 43$\pm$4 & 110$\pm$11 & 43.4$\pm$0.007 & 114$\pm$3\\
2021 Jun 5 & 59370 & SDSS 2.5m/BOSS & 2 & 4140$\pm$110 & 369$\pm$7 & 791$\pm$16 & 5900$\pm$430 & 58$\pm$3 & 159$\pm$8 & 43.4$\pm$0.005 & $121^{+2}_{-3}$\\ \hline
2021 Aug 17 & 59443 & FTN 2m/FLOYDS &2 & 3920$\pm$390 & 247$\pm$17 & 1248$\pm$87 & 4100$\pm$1700 & 32$\pm$7 & 204$\pm$42 & 43.8$\pm$0.009 & $130^{+10}_{-14}$ & ``Intermediate'' \\
2022 Jan 22 & 59601 & FTN 2m/FLOYDS &2 & 3200$\pm$460 & 269$\pm$44 & 1070$\pm$170 & 4900$\pm$3600 & 36$\pm$13 & 217$\pm$81 & 43.8$\pm$0.01 & $120^{+27}_{-24}$\\
2022 Feb 26\tablenotemark{*} & 59636 & ARC 3.5m/KOSMOS & 2 & 3720$\pm$120 & 298$\pm$13 & & 8870$\pm$460 & 102$\pm$4 & &  &\\
2022 Mar 25 & 59663 & FTN 2m/FLOYDS &2 & 4040$\pm$150 & 321$\pm$13 & 1243$\pm$51 & 4500$\pm$300 & 58$\pm$3 & 332$\pm$16 & 43.8$\pm$0.006 & $110^{+5}_{-6}$\\
2022 May 9\tablenotemark{*} & 59708 & SDSS 2.5m/BOSS &2& 4380$\pm$370 & 330$\pm$30 & & 4720$\pm$590 & 55$\pm$3 & &  &\\
2022 May 10\tablenotemark{*} & 59709 & SDSS 2.5m/BOSS &2& 4170$\pm$280 & 397$\pm$24 & & 4970$\pm$40 & 71$\pm$4 & &  &\\
2022 May 21\tablenotemark{*} & 59720 & HET 10m/LRS-2 & \tablenotemark{b} & 3680$\pm$280 & 299$\pm$44 & & 4820$\pm$350 & 61$\pm$3 & &  &\\
\enddata

\tablenotetext{a}{Aperture size: slit width or fiber diameter.}
\tablenotetext{b}{The spectrum was extracted from the integral field unit data cube after adopting a Gaussian point-spread function with a FWHM of $2\,.\!\!^{\prime\prime}1$. Approximately 75\% of the light is enclosed in an aperture of diameter equal to the FWHM and 94\% of the light is enclosed in an aperture of diameter of $2\times$FWHM.}
\tablenotetext{*}{These late spectra have large absolute flux calibration uncertainties and are not used for our main analysis. The corresponding FWHM and EW measurements are robust.}
\end{deluxetable*}

%%%%%%%%%%%%%%%%%%%%%%%%%%%%%%%%%%%%%%%%%%%%%%%%%%%%%%%%%%%%%%%%

\section{Discussion}
\label{sec:dicussion}

The dramatic variability events observed in \target\ are intriguing because the source seems to transit between relatively well-defined states, over merely a few months---among the fastest CLAGN transitions seen to date \cite[e.g.,][]{Guo16,Trakhtenbrot19,Ross20}.
These events can be interpreted in several ways: (1) an intrinsic change in the emission from the central engine and from the BLR due to a variable accretion flow, (2) an apparent change driven by variable obscuration along the \los, or (3) a more nuanced combination of the two, related to dust formation and sublimation.
We discuss these possibilities below.

\subsection{Variable Accretion Flow}
\label{subsec:var_acc}

To test the possibility of a change to the accretion flow, we consider the timescales associated with continuum variability and with the BLR response to it. 

As discussed in detail in many previous studies \cite[e.g.,][]{LaMassa15,Stern18}, the relevant timescales for drastic changes in thin, stable  accretion disks are thought to be much longer than what is observed in CLAGN transitions, even for the case of years-timescale transitions.
Even more sophisticated models do not achieve coherent disk transitions on intrayear timescales \cite[see detailed discussion in, e.g.,][and references therein]{Shen21}. 
In addition, while novel models involving magnetic flux inversion can explain rapid UV-optical flares \cite[][]{Scepi21}, their relevance to the dip seen in \target\ is not yet clear.
\target\ challenges common theoretical expectations even further, given the detailed nature of its rapid (months-timescale) transitions. 
Following the common calculations for other CLAGN, we note that the disk material infall timescale for \target\ is $\approx$80 yr \cite[following Eq.~5 in][]{LaMassa15}.
Observationally, however, changes to the accretion flow on timescales of several months have been inferred for \emph{some} CLAGNs events \cite[e.g.,][]{Gezari17,Trakhtenbrot19}.
We thus cannot exclude an accretion-driven mechanism to explain \target\ based solely on timescale arguments.

As for the BLR response, its timescale is expected to be dominated by the light-travel time between the ionizing source and the BLR, $t_{\rm{lt}}=\RBLR/c$, where \RBLR\ is the characteristic radius of the BLR.
Using the $\RBLR-L$ relation of \cite{Bentz13} and the measured \Lop\ we infer $\RBLR\simeq36$ and $16$ \ld, for the brightest (SDSS; 2001) and faintest (SDSS-V; 2021) spectra, respectively.
The corresponding light-travel timescales could thus be accommodated within the observed $\approx$60 days timescale of BLR variability of \target. 

Another expectation that arises for BLR clouds moving at Keplerian velocities that are responding to continuum variations is that the typical BEL velocities would follow ${\rm FWHM}\sim R^{-1/2}_{\rm{BLR}}\sim L^{-1/4}$ \cite[e.g.,][]{Barth15,Wang20}. 
Between the 2001 bright state and the 2016 intermediate state, \fwha\ \emph{decreased} by a factor of $1.34\pm0.05$ and \fwhb\ remained essentially constant (increased by a factor of $1.06\pm0.05$), while the expectation based on the luminosity--scaling relation is for an \emph{increase} by a factor of $\approx1.2$. 
Between the 2016 intermediate state and 2021 dim state, \fwha\ and \fwhb\ both increased, by factors of $1.12\pm0.04$ and $1.3\pm0.1$, respectively, consistent with a similar expectation of a uniform increase by a factor of $\approx1.2$. 

One significant challenge of the variable accretion flow hypothesis is to account for the observed spectral ratio between the intermediate and dim states, shown in Fig.~\ref{fig:ratio}. 
The lack of BLR-like features in the spectral ratio
means that, for each of the Balmer BELs, the dimming factor is comparable to the one of the adjacent continua. 
This result is especially surprising considering that 
the optical continuum is emitted from the outer parts of the accretion disk,
while the broad-line emission is fundamentally driven by the ionizing ($>$13.6 eV) radiation, which in turn originates from the inner parts of the disk.
While certainly these two forms of radiation are physically linked, there is no {\it a priori} reason to expect that they would scale linearly, as supported by the observed ``bluer-when-brighter'' trend \cite[see, e.g.,][and references therein]{Rumbaugh18}.
Specifically for \target, theoretical thin-disk spectral energy distributions that are calculated based on the observed \mbh\ and accretion rates for the intermediate--dim transition imply a variation in $L({>}13.6\,{\rm eV})$ by a factor of ${\approx}3.6{\times}$, compared to the observed factor of only ${\approx}2.4{\times}$ in the optical regime. 
A similar analysis for the bright-to-dim transition implies a variation by a factor of ${\approx}9{\times}$ in $L({>}13.6\,{\rm eV})$, and by a factor of ${\approx}5{\times}$ in the optical regime.
Even if some of the optical continuum originates from reprocessed UV light in the disk or as diffuse continuum from the BLR \cite[e.g.,][]{Chelouche19}, this cannot fully explain the interlinked continuum and line variations we observe.

Finally---and most importantly---it is challenging to explain how the variable accretion flow scenario would produce the well-defined dip seen in \target\ during 2021, and the (spectral) recovery back to a state essentially identical to the 2016 one.

\subsection{Variation Caused by a Crossing Cloud}
\label{subsec:var_obsc}

A variable obscuration scenario invokes an obscuring cloud intercepting the \los\ at the end of 2020, and then going out of the \los\ between 2021 June and August (Fig.~\ref{fig:photometry}). 
Another obscuring cloud, which entered our \los\ circa 2011, may explain the more gradual dimming between 2011 and 2013.
This scenario is supported by the ``dip'' in the 2021 ZTF optical light curve, and would naturally explain the similarity of the 2016 and most recent (LCO) spectra. 

\begin{figure*}
    \centering
    \includegraphics[width=1.0\textwidth]{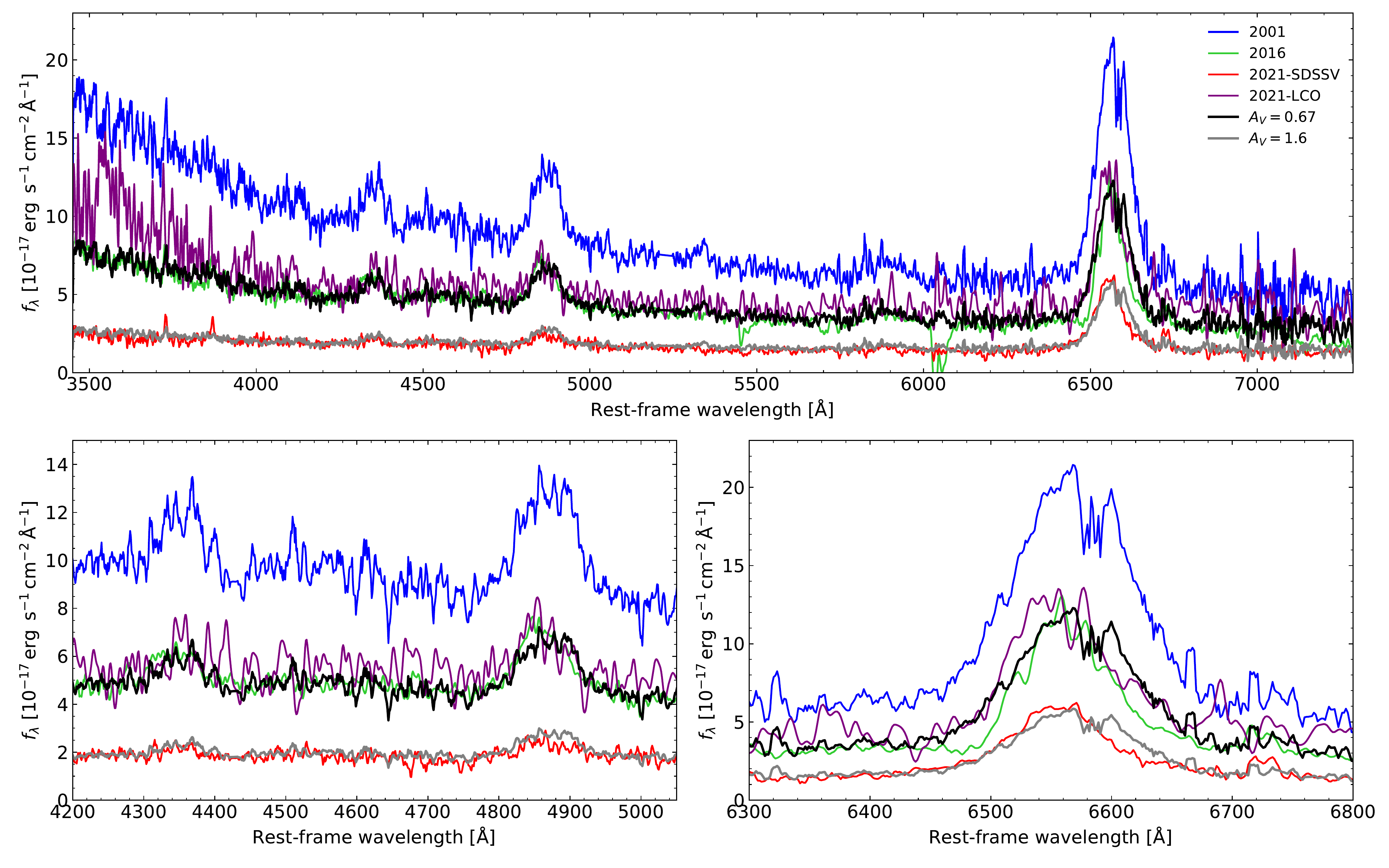}
    \caption{Representative host- and narrow-line-subtracted spectra of \target: 2001 bright state (blue), 2016 intermediate state (green), 2021 dim state (red; SDSS-V), and 2021 intermediate state (purple; LCO). Along those, we plot reddened spectra of the bright state, obtained by applying the ``Maiolino AGN'' extinction law. The appropriate $A_V$ values were selected to best reproduce the intermediate ($A_V=0.67$; black) and dim ($A_V=1.6$; gray) states.}
    \label{fig:extinction}
\end{figure*}

To quantitatively test the implications of this scenario, we applied the ``Maiolino AGN'' extinction law \cite[]{Maiolino01, Li07} to the 2001 bright-state spectrum. 
This relatively flat extinction law has been shown to be a good description of reddening in AGNs (see, e.g., \citealt{Maiolino01, Xie17}, and references therein, but also \citealt{Richards03}). 
For \target, this extinction law provides a good agreement between the various epochs. A similarly satisfactory agreement can also be achieved using a MW-like extinction law with $R_V=4.4$, or an LMC-like extinction law. 
The top panel of Figure~\ref{fig:extinction} shows representative host-galaxy- and narrow-line-subtracted observed spectra of \target, as well as those artificially reddened versions of the intermediate- and bright-state spectra that best match the dimmer states. 
Specifically, an intermediate-state spectrum  reddened by $A_V\approx0.89$ (black line) and/or a bright-state spectrum reddened by $A_V\approx1.6$ (light gray) match the dim-state spectrum remarkably well. 
In addition, a bright-state spectrum reddened by $A_V\approx0.67$ (dark gray) matches well the intermediate-state one.
Assuming a MW-like dust-to-gas ratio, this analysis would imply that a dust cloud with $N_{\rm H}\simeq1.6\times10^{21}\,\rm{cm}^{-2}$ has entered and exited our \los\ during the 2021 intermediate--dim--intermediate transition. Similarly, the bright-to-intermediate and bright-to-dim transitions would imply column densities of $N_{\rm{H}}=1.2$ and $2.8\times10^{21}\,\rm{cm}^{-2}$, respectively.\footnote{Note, however, that many AGNs are observed to have higher gas-to-dust ratios \cite[e.g.,][]{Maiolino01}.} 
The bottom panels focus on the broad Balmer emission lines (\Ha, \Hb, \Hg).
The reddened spectra of the bright state show a striking similarity to the intermediate- and dim-state spectra, across the entire observed wavelength range, which strongly supports the variable obscuration interpretation. These similarities are further demonstrated by the nearly uniform variations in line and continuum emission seen in Fig.~\ref{fig:ratio}, which also shows how the several extinction curves considered here can reasonably explain the observed intermediate-to-dim spectral ratio.

The main challenge to the transient obscuration scenario relates to the observed and expected timescales. 
Specifically, assuming an obscurer moving at Keplerian velocities within the black hole sphere of influence, the appropriate dynamical timescale is $\tdyn\simeq\left(\Rcloud^3/G\mbh\right)^{1/2}$, where \Rcloud is the distance of the obscuring cloud from the SMBH. 
Plugging in the minimal value of $\Rcloud=\RBLR$ produces $\tdyn\approx6\,\rm{yr}$, a factor of $\sim$35 longer than the observed shortest variability timescale of \target.
A more detailed derivation that accounts for the projected \los\ motion of the obscurer, following \citet[][see their Eq.~4]{LaMassa15}, yields yet longer crossing timescales of $\gtrsim18$ yr. 
This argument becomes particularly relevant if the obscurer is part of the (clumpy) torus, i.e. if $\Rcloud= {\rm few}\times\RBLR$ \cite[e.g.,][and references therein]{Risaliti02,Koshida14}. 
More generally, for an obscuring object in the AGN host galaxy to ``reveal'' the BLR, the obscurer has to travel a physical distance of $\RBLR\gtrsim30$ \ld\ within the $<$2 months span of the fastest transition observed, and thus must move at an unreasonably high (tangential) velocity of $\gtrsim0.5c$. 
Furthermore, an obscuring cloud that covers a substantial portion of the BLR would have to be of (at least) a comparable size (i.e., ${\sim}\RBLR$), which is larger by at least an order of magnitude than the expected size for clouds in the torus \cite[]{Elitzur08}.
We note that the optical variability in the spectra of some Seyfert 1.8 and 1.9 galaxies has been reported to be best explained by variable obscuration \cite[e.g.,][]{Goodrich95}. Those transitions, however, were identified using spectra taken several years apart, and were thus consistent with much slower obscurer velocities, of $\sim100-1000\,\kms$. 
In addition, the presence of obscuring clouds moving in and out of the \los\ toward some AGNs were identified by several X-ray studies \cite[e.g.,][]{Risaliti05, Maiolino10, Markowitz14}. However, in such cases the obscuring clouds must cover only the much more compact X-ray emitting region, and can thus move at lower velocities and have a smaller physical size, compared to the obscurer discussed here. In fact, the obscurers in X-ray studies are suspected to be BLR clouds \cite[e.g.,][]{Maiolino10}, while the obscurer in the case of \target\ cannot be a BLR cloud, as it is required to obscure a large part of the BLR itself.

In principle, the challenges posed by the BLR-obscuring cloud velocity and size could be addressed by a very compact BLR, or a configuration where only a small part of the BLR is visible in our \los, as $v_{\rm cloud}\propto \Rcloud \propto \RBLR$. 
We note, however, that the observed broad \Ha strength (i.e., $L$[b\Ha]) does not support a particularly small BLR or one with a low covering factor (see distributions in, e.g., \citealt{GreeneHo07} or \citealt{SternLaor12} for comparison).

An additional challenge to the variable obscuration interpretation is posed by the WISE photometry (Figure~\ref{fig:photometry}), which shows a clear dimming of $\approx0.4$ mag in the $W1$ and $W2$ bands sometime between 2011 and 2014---as expected for the reverberation response from the dusty torus of the UV/optical accretion disk continuum variability \cite[e.g.,][]{Yang20}.
The WISE light curve also shows hints of a weaker dimming in the most recent data point, in 2021 July, coincident with the recovery seen in the optical (ZTF) light curve. 
The interpretation of this second WISE dimming is limited by large uncertainties and the low cadence of measurements ($\approx$6 months). 
If we assume that the dimmest WISE measurement is indeed associated with the rapid 2021 optical dip, it would be inconsistent with a variable obscuration scenario, as the WISE bands are not expected to be affected by (dust) obscuration. 
Thus, the limited WISE data in hand favor variable accretion as an explanation for the slow 2011 dimming, and perhaps also for the rapid 2021 event.

\subsection{Rapid Dust Formation and Sublimation}

The rapid changes in \target, combined with the similarity between the extinction-reddened ``bright-state'' spectrum and the fainter spectra, raise the possibility of variable obscuration occurring on the BLR light-travel timescale, rather than dynamical timescale. 
In such a scenario, the rapid 2021 changes may be driven by a temporary decrease in the UV emission from the central engine, which allowed the formation of dust in the previously sublimated, innermost parts of the dusty torus. After the UV emission reverted to its earlier state, the dust sublimated to its previous levels.
Since the innermost part of the torus is thought to reside just outside the (dust-free) BLR \cite[e.g.,][]{Risaliti02, Koshida14}, the time required for the enhanced radiation to reach the torus is only slightly longer than the BLR light-crossing time, i.e., $\rm{few}\times t_{\rm lt, BLR}$, which is consistent with the fastest variability seen in \target. 
The dust formation and sublimation timescales themselves are sufficiently short for BLR-like gas densities (i.e., up to $\sim$1 month  for $n \simeq10^9\,\rm{cm}^{-3}$; e.g., \citealt{Draine09, BL18}).\footnote{However, for lower gas densities the timescales could become challengingly long, as $t\propto n^{-1}$ \cite[e.g.,][]{Draine09}.}

A key challenge for this scenario is that it necessitates a considerable and rapid change in the UV continuum emission that did not significantly and directly affect the observed optical continuum and broad-line emission, as the spectral variations are fully accounted for by varying obscuration effects (Fig.~\ref{fig:extinction}). 
This can be explained if the changes in the dust-forming UV drop and its subsequent, direct effect on the BLR emission  
occurred at a time unrelated to the observed spectroscopic changes. 
However, we cannot find direct evidence to support such a nuanced, highly interlinked, and somewhat contrived scenario within the data in hand.
Moreover, the dust formation and sublimation scenario fails to explain the 2021 recovery of \target\ to a spectral state essentially indistinguishable from the predip one, in terms of both continuum and broad-line emission (similarly to the variable accretion scenario discussed in Section \ref{subsec:var_acc}).
All this renders the dust formation and sublimation scenario rather challenging for explaining the rapid dimming and recovery observed during 2021.

%%%%%%%%%%%%%%%%%%%%%%%%%%%%%%%%%%%%%%%%%%%%%%%%%%%%%%%%%%%%%%%%
\section{Conclusions}
\label{sec:summary}

We presented multiepoch optical spectroscopy of \target, a ``changing-look'' AGN identified in the recently initiated SDSS-V project. \target\ was observed in three rather distinct spectral states, with the fastest transition occurring within a timescale of less than 2 months---one of the fastest significant spectral changes observed to date.
 
We explored several possible explanations for the observed extreme variability of \target, driven by variations in either the accretion flow and/or the \los\ obscuration.
The observed variability timescale appears to be consistent with changes in the radiation from the accretion flow impinging on the BLR, and also with observations of other CLAGN events driven by accretion variations (Section \ref{subsec:var_acc}). This explanation is further supported by the dimming in the WISE light curve between 2011 and 2014 (Fig.~\ref{fig:photometry}).

Nonetheless, the excellent agreement between the observed dimmer spectra and the (artificially) dust-reddened bright spectrum is more naturally explained by an obscuration-driven scenario (Fig.~\ref{fig:extinction}). 
Moreover---and perhaps most importantly---the 2021 ``dip'' in the light curve of \target\ (Fig.~\ref{fig:photometry}), which is also associated with a temporary transition to a dimmer and redder spectral state, followed by a recovery to a state essentially identical to that preceding the dip, is exactly what one would expect from an event driven by variable \los\ obscuration.
This explanation, however, implies exceptionally high velocity and large size for the obscurer, or else a challengingly compact BLR (Section \ref{subsec:var_obsc}).

The transitions observed in \target\ 
demonstrate the richness, potential, and challenges presented by CLAGNs, as well as 
the importance of intrayear cadence spectroscopy for large AGN samples.
The discovery of this unusual event is an early result of the SDSS-V Black Hole Mapper science program - which is
spectroscopically monitoring tens of 1000s of AGNs on timescales of days to years, and is thus poised to discover, monitor, and survey many more systems of this sort. 
Responsive, multiwavelength follow-up observations can greatly help to determine the physical drivers of these events.

%%%%%%%%%%%%%%%%%%%%%%%%%%%%%%%%%%%%%%%%%%%%%%%%%%%%%%%%%%%%%%%%
\begin{acknowledgments}

We thank Aigen Li for providing us the AGN-related extinction law data (inspired by the \citealt{Li07} review).

G.Z. and B.T. acknowledge support from the European Research Council (ERC) under the European Union's Horizon 2020 research and innovation program (grant agreement 950533) and from the Israel Science Foundation (grant 1849/19).
J.R.T. acknowledges support from NSF grants CAREER-1945546, AST-2009539, and AST-2108668.
We acknowledge funding from ANID - Millennium Science Initiative Program - ICN12\_009 (F.E.B., L.H.G.), CATA-Basal - FB210003 (F.E.B., R.J.A.), and FONDECYT Regular - 1190818 (F.E.B.) and 1200495 (F.E.B.).
C.R. acknowledges support from the Fondecyt Iniciacion grant 11190831 and ANID BASAL project FB210003.
R.J.A. was also supported by FONDECYT grant No. 1191124. 
D.H. is supported by DLR grant FKZ 50OR2003.
M.K. acknowledges support from DFG grant KR 3338/4-1.
% and by ANID BASAL project FB210003.
X.L. acknowledges support by NSF grants AST-2108162 and AST-2206499.
M. L. M.-A. acknowledges financial support from Millenium Nucleus NCN$19\_058$ (TITANs).
H.I.M. acknowledges a support grant from the Joint Committee ESO-Government of Chile (ORP 028/2020).
%
% Jamison Burke:
J.B. is supported by NSF grants AST-1911151 and AST-1911225, as well as by NASA grant 80NSSC19kf1639.

Funding for the Sloan Digital Sky Survey V has been provided by the Alfred P. Sloan Foundation, the Heising-Simons Foundation, and the Participating Institutions. SDSS acknowledges support and resources from the Center for High-Performance Computing at the University of Utah. The SDSS website is \url{www.sdss5.org}.

SDSS is managed by the Astrophysical Research Consortium for the Participating Institutions of the SDSS Collaboration, including the Carnegie Institution for Science, Chilean National Time Allocation Committee (CNTAC) ratified researchers, the Gotham Participation Group, Harvard University, Heidelberg University, The Johns Hopkins University, L'Ecole polytechnique f\'{e}d\'{e}rale de Lausanne (EPFL), Leibniz-Institut f\"{u}r Astrophysik Potsdam (AIP), Max-Planck-Institut f\"{u}r Astronomie (MPIA Heidelberg), Max-Planck-Institut f\"{u}r Extraterrestrische Physik (MPE), Nanjing University, National Astronomical Observatories of China (NAOC), New Mexico State University, The Ohio State University, Pennsylvania State University, Smithsonian Astrophysical Observatory, Space Telescope Science Institute (STScI), the Stellar Astrophysics Participation Group, Universidad Nacional Aut\'{o}noma de M\'{e}xico, University of Arizona, University of Colorado Boulder, University of Illinois at Urbana-Champaign, University of Toronto, University of Utah, University of Virginia, Yale University, and Yunnan University.

The ZTF forced-photometry service was funded under the Heising-Simons Foundation grant No. 12540303 (PI: Graham). The Hobby--Eberly Telescope (HET) is a joint project of the University of Texas at Austin, the Pennsylvania State University, Ludwig-Maximillians-Universitaet Muenchen, and Georg-August Universitaet Goettingen. The HET is named in honor of its principal benefactors, William P. Hobby and Robert E. Eberly.

\end{acknowledgments}

\facilities{Sloan (SDSS and BOSS), ING:Herschel (ISIS), LCOGT (FTN: FLOYDS), ARC (KOSMOS), HET (LRS-2).}

\software{{\tt AstroPy} \citep{Astropy18},
{\tt Matplotlib} \citep{Matplotlib07}, 
{\tt NumPy} \citep{NumPy20}, {\tt SciPy} \citep{SciPy20}, {\tt PyQSOFit} \citep{QSOFit, Shen19}}

% \smallskip
\clearpage
\bibliography{J1628+4329}{}

\begin{thebibliography}{}
\expandafter\ifx\csname natexlab\endcsname\relax\def\natexlab#1{#1}\fi
\providecommand{\url}[1]{\href{#1}{#1}}
\providecommand{\dodoi}[1]{doi:~\href{http://doi.org/#1}{\nolinkurl{#1}}}
\providecommand{\doeprint}[1]{\href{http://ascl.net/#1}{\nolinkurl{http://ascl.net/#1}}}
\providecommand{\doarXiv}[1]{\href{https://arxiv.org/abs/#1}{\nolinkurl{https://arxiv.org/abs/#1}}}

\bibitem[{{Astropy Collaboration} {et~al.}(2018){Astropy Collaboration},
  {Price-Whelan}, {Sip{\H{o}}cz}, {G{\"u}nther}, {Lim}, {Crawford}, {Conseil},
  {Shupe}, {Craig}, {Dencheva}, {Ginsburg}, {VanderPlas}, {Bradley},
  {P{\'e}rez-Su{\'a}rez}, {de Val-Borro}, {Aldcroft}, {Cruz}, {Robitaille},
  {Tollerud}, {Ardelean}, {Babej}, {Bach}, {Bachetti}, {Bakanov}, {Bamford},
  {Barentsen}, {Barmby}, {Baumbach}, {Berry}, {Biscani}, {Boquien}, {Bostroem},
  {Bouma}, {Brammer}, {Bray}, {Breytenbach}, {Buddelmeijer}, {Burke},
  {Calderone}, {Cano Rodr{\'\i}guez}, {Cara}, {Cardoso}, {Cheedella}, {Copin},
  {Corrales}, {Crichton}, {D'Avella}, {Deil}, {Depagne}, {Dietrich}, {Donath},
  {Droettboom}, {Earl}, {Erben}, {Fabbro}, {Ferreira}, {Finethy}, {Fox},
  {Garrison}, {Gibbons}, {Goldstein}, {Gommers}, {Greco}, {Greenfield},
  {Groener}, {Grollier}, {Hagen}, {Hirst}, {Homeier}, {Horton}, {Hosseinzadeh},
  {Hu}, {Hunkeler}, {Ivezi{\'c}}, {Jain}, {Jenness}, {Kanarek}, {Kendrew},
  {Kern}, {Kerzendorf}, {Khvalko}, {King}, {Kirkby}, {Kulkarni}, {Kumar},
  {Lee}, {Lenz}, {Littlefair}, {Ma}, {Macleod}, {Mastropietro}, {McCully},
  {Montagnac}, {Morris}, {Mueller}, {Mumford}, {Muna}, {Murphy}, {Nelson},
  {Nguyen}, {Ninan}, {N{\"o}the}, {Ogaz}, {Oh}, {Parejko}, {Parley}, {Pascual},
  {Patil}, {Patil}, {Plunkett}, {Prochaska}, {Rastogi}, {Reddy Janga},
  {Sabater}, {Sakurikar}, {Seifert}, {Sherbert}, {Sherwood-Taylor}, {Shih},
  {Sick}, {Silbiger}, {Singanamalla}, {Singer}, {Sladen}, {Sooley},
  {Sornarajah}, {Streicher}, {Teuben}, {Thomas}, {Tremblay}, {Turner},
  {Terr{\'o}n}, {van Kerkwijk}, {de la Vega}, {Watkins}, {Weaver}, {Whitmore},
  {Woillez}, {Zabalza}, \& {Astropy Contributors}}]{Astropy18}
{Astropy Collaboration}, {Price-Whelan}, A.~M., {Sip{\H{o}}cz}, B.~M., {et~al.}
  2018, \aj, 156, 123, \dodoi{10.3847/1538-3881/aabc4f}

\bibitem[{{Barth} {et~al.}(2015){Barth}, {Bennert}, {Canalizo}, {Filippenko},
  {Gates}, {Greene}, {Li}, {Malkan}, {Pancoast}, {Sand}, {Stern}, {Treu},
  {Woo}, {Assef}, {Bae}, {Brewer}, {Cenko}, {Clubb}, {Cooper},
  {Diamond-Stanic}, {Hiner}, {H{\"o}nig}, {Hsiao}, {Kandrashoff}, {Lazarova},
  {Nierenberg}, {Rex}, {Silverman}, {Tollerud}, \& {Walsh}}]{Barth15}
{Barth}, A.~J., {Bennert}, V.~N., {Canalizo}, G., {et~al.} 2015, \apjs, 217,
  26, \dodoi{10.1088/0067-0049/217/2/26}

\bibitem[{{Baskin} \& {Laor}(2018)}]{BL18}
{Baskin}, A., \& {Laor}, A. 2018, \mnras, 474, 1970,
  \dodoi{10.1093/mnras/stx2850}

\bibitem[{{Bentz} {et~al.}(2013){Bentz}, {Denney}, {Grier}, {Barth},
  {Peterson}, {Vestergaard}, {Bennert}, {Canalizo}, {De Rosa}, {Filippenko},
  {Gates}, {Greene}, {Li}, {Malkan}, {Pogge}, {Stern}, {Treu}, \&
  {Woo}}]{Bentz13}
{Bentz}, M.~C., {Denney}, K.~D., {Grier}, C.~J., {et~al.} 2013, \apj, 767, 149,
  \dodoi{10.1088/0004-637X/767/2/149}

\bibitem[{{Brown} {et~al.}(2013){Brown}, {Baliber}, {Bianco}, {Bowman},
  {Burleson}, {Conway}, {Crellin}, {Depagne}, {De Vera}, {Dilday}, {Dragomir},
  {Dubberley}, {Eastman}, {Elphick}, {Falarski}, {Foale}, {Ford}, {Fulton},
  {Garza}, {Gomez}, {Graham}, {Greene}, {Haldeman}, {Hawkins}, {Haworth},
  {Haynes}, {Hidas}, {Hjelstrom}, {Howell}, {Hygelund}, {Lister}, {Lobdill},
  {Martinez}, {Mullins}, {Norbury}, {Parrent}, {Paulson}, {Petry}, {Pickles},
  {Posner}, {Rosing}, {Ross}, {Sand}, {Saunders}, {Shobbrook}, {Shporer},
  {Street}, {Thomas}, {Tsapras}, {Tufts}, {Valenti}, {Vander Horst}, {Walker},
  {White}, \& {Willis}}]{Brown13}
{Brown}, T.~M., {Baliber}, N., {Bianco}, F.~B., {et~al.} 2013, \pasp, 125,
  1031, \dodoi{10.1086/673168}

\bibitem[{{Chambers} {et~al.}(2016){Chambers}, {Magnier}, {Metcalfe},
  {Flewelling}, {Huber}, {Waters}, {Denneau}, {Draper}, {Farrow}, {Finkbeiner},
  {Holmberg}, {Koppenhoefer}, {Price}, {Rest}, {Saglia}, {Schlafly}, {Smartt},
  {Sweeney}, {Wainscoat}, {Burgett}, {Chastel}, {Grav}, {Heasley}, {Hodapp},
  {Jedicke}, {Kaiser}, {Kudritzki}, {Luppino}, {Lupton}, {Monet}, {Morgan},
  {Onaka}, {Shiao}, {Stubbs}, {Tonry}, {White}, {Ba{\~n}ados}, {Bell},
  {Bender}, {Bernard}, {Boegner}, {Boffi}, {Botticella}, {Calamida},
  {Casertano}, {Chen}, {Chen}, {Cole}, {Deacon}, {Frenk}, {Fitzsimmons},
  {Gezari}, {Gibbs}, {Goessl}, {Goggia}, {Gourgue}, {Goldman}, {Grant},
  {Grebel}, {Hambly}, {Hasinger}, {Heavens}, {Heckman}, {Henderson}, {Henning},
  {Holman}, {Hopp}, {Ip}, {Isani}, {Jackson}, {Keyes}, {Koekemoer}, {Kotak},
  {Le}, {Liska}, {Long}, {Lucey}, {Liu}, {Martin}, {Masci}, {McLean}, {Mindel},
  {Misra}, {Morganson}, {Murphy}, {Obaika}, {Narayan}, {Nieto-Santisteban},
  {Norberg}, {Peacock}, {Pier}, {Postman}, {Primak}, {Rae}, {Rai}, {Riess},
  {Riffeser}, {Rix}, {R{\"o}ser}, {Russel}, {Rutz}, {Schilbach}, {Schultz},
  {Scolnic}, {Strolger}, {Szalay}, {Seitz}, {Small}, {Smith}, {Soderblom},
  {Taylor}, {Thomson}, {Taylor}, {Thakar}, {Thiel}, {Thilker}, {Unger},
  {Urata}, {Valenti}, {Wagner}, {Walder}, {Walter}, {Watters}, {Werner},
  {Wood-Vasey}, \& {Wyse}}]{Chambers16}
{Chambers}, K.~C., {Magnier}, E.~A., {Metcalfe}, N., {et~al.} 2016, arXiv
  e-prints, arXiv:1612.05560.
\newblock \doarXiv{1612.05560}

\bibitem[{{Chelouche} {et~al.}(2019){Chelouche}, {Pozo Nu{\~n}ez}, \&
  {Kaspi}}]{Chelouche19}
{Chelouche}, D., {Pozo Nu{\~n}ez}, F., \& {Kaspi}, S. 2019, Nature Astronomy,
  3, 251, \dodoi{10.1038/s41550-018-0659-x}

\bibitem[{{Draine}(2009)}]{Draine09}
{Draine}, B.~T. 2009, in Astronomical Society of the Pacific Conference Series,
  Vol. 414, Cosmic Dust - Near and Far, ed. T.~{Henning}, E.~{Gr{\"u}n}, \&
  J.~{Steinacker}, 453.
\newblock \doarXiv{0903.1658}

\bibitem[{{Elitzur}(2008)}]{Elitzur08}
{Elitzur}, M. 2008, \nar, 52, 274, \dodoi{10.1016/j.newar.2008.06.010}

\bibitem[{{Flewelling} {et~al.}(2020){Flewelling}, {Magnier}, {Chambers},
  {Heasley}, {Holmberg}, {Huber}, {Sweeney}, {Waters}, {Calamida}, {Casertano},
  {Chen}, {Farrow}, {Hasinger}, {Henderson}, {Long}, {Metcalfe}, {Narayan},
  {Nieto-Santisteban}, {Norberg}, {Rest}, {Saglia}, {Szalay}, {Thakar},
  {Tonry}, {Valenti}, {Werner}, {White}, {Denneau}, {Draper}, {Hodapp},
  {Jedicke}, {Kaiser}, {Kudritzki}, {Price}, {Wainscoat}, {Chastel}, {McLean},
  {Postman}, \& {Shiao}}]{Flewelling20}
{Flewelling}, H.~A., {Magnier}, E.~A., {Chambers}, K.~C., {et~al.} 2020, \apjs,
  251, 7, \dodoi{10.3847/1538-4365/abb82d}

\bibitem[{{F{\"o}rster} {et~al.}(2021){F{\"o}rster}, {Cabrera-Vives},
  {Castillo-Navarrete}, {Est{\'e}vez}, {S{\'a}nchez-S{\'a}ez}, {Arredondo},
  {Bauer}, {Carrasco-Davis}, {Catelan}, {Elorrieta}, {Eyheramendy}, {Huijse},
  {Pignata}, {Reyes}, {Reyes}, {Rodr{\'\i}guez-Mancini}, {Ruz-Mieres},
  {Valenzuela}, {{\'A}lvarez-Maldonado}, {Astorga}, {Borissova}, {Clocchiatti},
  {De Cicco}, {Donoso-Oliva}, {Hern{\'a}ndez-Garc{\'\i}a}, {Graham},
  {Jord{\'a}n}, {Kurtev}, {Mahabal}, {Maureira}, {Mu{\~n}oz-Arancibia},
  {Molina-Ferreiro}, {Moya}, {Palma}, {P{\'e}rez-Carrasco}, {Protopapas},
  {Romero}, {Sabatini-Gacitua}, {S{\'a}nchez}, {San Mart{\'\i}n},
  {Sep{\'u}lveda-Cobo}, {Vera}, \& {Vergara}}]{Forster21}
{F{\"o}rster}, F., {Cabrera-Vives}, G., {Castillo-Navarrete}, E., {et~al.}
  2021, \aj, 161, 242, \dodoi{10.3847/1538-3881/abe9bc}

\bibitem[{{Frederick} {et~al.}(2019){Frederick}, {Gezari}, {Graham}, {Cenko},
  {van Velzen}, {Stern}, {Blagorodnova}, {Kulkarni}, {Yan}, {De}, {Fremling},
  {Hung}, {Kara}, {Shupe}, {Ward}, {Bellm}, {Dekany}, {Duev}, {Feindt},
  {Giomi}, {Kupfer}, {Laher}, {Masci}, {Miller}, {Neill}, {Ngeow}, {Patterson},
  {Porter}, {Rusholme}, {Sollerman}, \& {Walters}}]{Frederick19}
{Frederick}, S., {Gezari}, S., {Graham}, M.~J., {et~al.} 2019, \apj, 883, 31,
  \dodoi{10.3847/1538-4357/ab3a38}

\bibitem[{{Gaskell} \& {Harrington}(2018)}]{GH18}
{Gaskell}, C.~M., \& {Harrington}, P.~Z. 2018, \mnras, 478, 1660,
  \dodoi{10.1093/mnras/sty848}

\bibitem[{{Gezari} {et~al.}(2017){Gezari}, {Hung}, {Cenko}, {Blagorodnova},
  {Yan}, {Kulkarni}, {Mooley}, {Kong}, {Cantwell}, {Yu}, {Cao}, {Fremling},
  {Neill}, {Ngeow}, {Nugent}, \& {Wozniak}}]{Gezari17}
{Gezari}, S., {Hung}, T., {Cenko}, S.~B., {et~al.} 2017, \apj, 835, 144,
  \dodoi{10.3847/1538-4357/835/2/144}

\bibitem[{{Goodrich}(1995)}]{Goodrich95}
{Goodrich}, R.~W. 1995, \apj, 440, 141, \dodoi{10.1086/175256}

\bibitem[{{Green} {et~al.}(2022){Green}, {Pulgarin-Duque}, {Anderson},
  {MacLeod}, {Eracleous}, {Ruan}, {Runnoe}, {Graham}, {Roulston}, {Schneider},
  {Ahlf}, {Bizyaev}, {Brownstein}, {del Casal}, {Dodd}, {Hoover}, {Matt},
  {Merloni}, {Pan}, {Ramirez}, \& {Ridder}}]{Green22}
{Green}, P.~J., {Pulgarin-Duque}, L., {Anderson}, S.~F., {et~al.} 2022, \apj,
  933, 180, \dodoi{10.3847/1538-4357/ac743f}

\bibitem[{{Greene} \& {Ho}(2007)}]{GreeneHo07}
{Greene}, J.~E., \& {Ho}, L.~C. 2007, \apj, 667, 131, \dodoi{10.1086/520497}

\bibitem[{{Gunn} {et~al.}(2006){Gunn}, {Siegmund}, {Mannery}, {Owen}, {Hull},
  {Leger}, {Carey}, {Knapp}, {York}, {Boroski}, {Kent}, {Lupton}, {Rockosi},
  {Evans}, {Waddell}, {Anderson}, {Annis}, {Barentine}, {Bartoszek}, {Bastian},
  {Bracker}, {Brewington}, {Briegel}, {Brinkmann}, {Brown}, {Carr},
  {Czarapata}, {Drennan}, {Dombeck}, {Federwitz}, {Gillespie}, {Gonzales},
  {Hansen}, {Harvanek}, {Hayes}, {Jordan}, {Kinney}, {Klaene}, {Kleinman},
  {Kron}, {Kresinski}, {Lee}, {Limmongkol}, {Lindenmeyer}, {Long}, {Loomis},
  {McGehee}, {Mantsch}, {Neilsen}, {Neswold}, {Newman}, {Nitta}, {Peoples},
  {Pier}, {Prieto}, {Prosapio}, {Rivetta}, {Schneider}, {Snedden}, \&
  {Wang}}]{Gunn06}
{Gunn}, J.~E., {Siegmund}, W.~A., {Mannery}, E.~J., {et~al.} 2006, \aj, 131,
  2332, \dodoi{10.1086/500975}

\bibitem[{{Guo} {et~al.}(2018){Guo}, {Shen}, \& {Wang}}]{QSOFit}
{Guo}, H., {Shen}, Y., \& {Wang}, S. 2018, {PyQSOFit: Python code to fit the
  spectrum of quasars}, Astrophysics Source Code Library.
\newblock \doeprint{1809.008}

\bibitem[{{Guo} {et~al.}(2016){Guo}, {Malkan}, {Gu}, {Li}, {Prochaska}, {Ma},
  {You}, {Zafar}, \& {Liao}}]{Guo16}
{Guo}, H., {Malkan}, M.~A., {Gu}, M., {et~al.} 2016, \apj, 826, 186,
  \dodoi{10.3847/0004-637X/826/2/186}

\bibitem[{{Guolo} {et~al.}(2021){Guolo}, {Ruschel-Dutra}, {Grupe}, {Peterson},
  {Storchi-Bergmann}, {Schimoia}, {Nemmen}, \& {Robinson}}]{Guolo21}
{Guolo}, M., {Ruschel-Dutra}, D., {Grupe}, D., {et~al.} 2021, \mnras, 508, 144,
  \dodoi{10.1093/mnras/stab2550}

\bibitem[{Harris {et~al.}(2020)Harris, Millman, van~der Walt, Gommers,
  Virtanen, Cournapeau, Wieser, Taylor, Berg, Smith, Kern, Picus, Hoyer, van
  Kerkwijk, Brett, Haldane, del R{\'{i}}o, Wiebe, Peterson,
  G{\'{e}}rard-Marchant, Sheppard, Reddy, Weckesser, Abbasi, Gohlke, \&
  Oliphant}]{NumPy20}
Harris, C.~R., Millman, K.~J., van~der Walt, S.~J., {et~al.} 2020, Nature, 585,
  357, \dodoi{10.1038/s41586-020-2649-2}

\bibitem[{{Hern{\'a}ndez-Garc{\'\i}a}
  {et~al.}(2017){Hern{\'a}ndez-Garc{\'\i}a}, {Masegosa},
  {Gonz{\'a}lez-Mart{\'\i}n}, {M{\'a}rquez}, {Guainazzi}, \&
  {Panessa}}]{Hernandez17}
{Hern{\'a}ndez-Garc{\'\i}a}, L., {Masegosa}, J., {Gonz{\'a}lez-Mart{\'\i}n},
  O., {et~al.} 2017, \aap, 602, A65, \dodoi{10.1051/0004-6361/201730476}

\bibitem[{{Hunter}(2007)}]{Matplotlib07}
{Hunter}, J.~D. 2007, Computing in Science and Engineering, 9, 90,
  \dodoi{10.1109/MCSE.2007.55}

\bibitem[{{Kollmeier} {et~al.}(2017){Kollmeier}, {Zasowski}, {Rix}, {Johns},
  {Anderson}, {Drory}, {Johnson}, {Pogge}, {Bird}, {Blanc}, {Brownstein},
  {Crane}, {De Lee}, {Klaene}, {Kreckel}, {MacDonald}, {Merloni}, {Ness},
  {O'Brien}, {Sanchez-Gallego}, {Sayres}, {Shen}, {Thakar}, {Tkachenko},
  {Aerts}, {Blanton}, {Eisenstein}, {Holtzman}, {Maoz}, {Nandra}, {Rockosi},
  {Weinberg}, {Bovy}, {Casey}, {Chaname}, {Clerc}, {Conroy}, {Eracleous},
  {G{\"a}nsicke}, {Hekker}, {Horne}, {Kauffmann}, {McQuinn}, {Pellegrini},
  {Schinnerer}, {Schlafly}, {Schwope}, {Seibert}, {Teske}, \& {van
  Saders}}]{Kollmeier17}
{Kollmeier}, J.~A., {Zasowski}, G., {Rix}, H.-W., {et~al.} 2017, arXiv
  e-prints, arXiv:1711.03234.
\newblock \doarXiv{1711.03234}

\bibitem[{{Koshida} {et~al.}(2014){Koshida}, {Minezaki}, {Yoshii}, {Kobayashi},
  {Sakata}, {Sugawara}, {Enya}, {Suganuma}, {Tomita}, {Aoki}, \&
  {Peterson}}]{Koshida14}
{Koshida}, S., {Minezaki}, T., {Yoshii}, Y., {et~al.} 2014, \apj, 788, 159,
  \dodoi{10.1088/0004-637X/788/2/159}

\bibitem[{{LaMassa} {et~al.}(2015){LaMassa}, {Cales}, {Moran}, {Myers},
  {Richards}, {Eracleous}, {Heckman}, {Gallo}, \& {Urry}}]{LaMassa15}
{LaMassa}, S.~M., {Cales}, S., {Moran}, E.~C., {et~al.} 2015, \apj, 800, 144,
  \dodoi{10.1088/0004-637X/800/2/144}

\bibitem[{{Li}(2007)}]{Li07}
{Li}, A. 2007, in Astronomical Society of the Pacific Conference Series, Vol.
  373, The Central Engine of Active Galactic Nuclei, ed. L.~C. {Ho} \& J.~W.
  {Wang}, 561

\bibitem[{{Liu} {et~al.}(2022){Liu}, {Luo}, {Brandt}, {Huang}, {Pu}, {Yi}, \&
  {Yu}}]{Liu22}
{Liu}, H., {Luo}, B., {Brandt}, W.~N., {et~al.} 2022, \apj, 930, 53,
  \dodoi{10.3847/1538-4357/ac6265}

\bibitem[{{MacLeod} {et~al.}(2019){MacLeod}, {Green}, {Anderson}, {Bruce},
  {Eracleous}, {Graham}, {Homan}, {Lawrence}, {LeBleu}, {Ross}, {Ruan},
  {Runnoe}, {Stern}, {Burgett}, {Chambers}, {Kaiser}, {Magnier}, \&
  {Metcalfe}}]{MacLeod19}
{MacLeod}, C.~L., {Green}, P.~J., {Anderson}, S.~F., {et~al.} 2019, \apj, 874,
  8, \dodoi{10.3847/1538-4357/ab05e2}

\bibitem[{{Maiolino} {et~al.}(2001){Maiolino}, {Marconi}, {Salvati},
  {Risaliti}, {Severgnini}, {Oliva}, {La Franca}, \& {Vanzi}}]{Maiolino01}
{Maiolino}, R., {Marconi}, A., {Salvati}, M., {et~al.} 2001, \aap, 365, 28,
  \dodoi{10.1051/0004-6361:20000177}

\bibitem[{{Maiolino} {et~al.}(2010){Maiolino}, {Risaliti}, {Salvati},
  {Pietrini}, {Torricelli-Ciamponi}, {Elvis}, {Fabbiano}, {Braito}, \&
  {Reeves}}]{Maiolino10}
{Maiolino}, R., {Risaliti}, G., {Salvati}, M., {et~al.} 2010, \aap, 517, A47,
  \dodoi{10.1051/0004-6361/200913985}

\bibitem[{{Markowitz} {et~al.}(2014){Markowitz}, {Krumpe}, \&
  {Nikutta}}]{Markowitz14}
{Markowitz}, A.~G., {Krumpe}, M., \& {Nikutta}, R. 2014, \mnras, 439, 1403,
  \dodoi{10.1093/mnras/stt2492}

\bibitem[{{Masci} {et~al.}(2019){Masci}, {Laher}, {Rusholme}, {Shupe}, {Groom},
  {Surace}, {Jackson}, {Monkewitz}, {Beck}, {Flynn}, {Terek}, {Landry},
  {Hacopians}, {Desai}, {Howell}, {Brooke}, {Imel}, {Wachter}, {Ye}, {Lin},
  {Cenko}, {Cunningham}, {Rebbapragada}, {Bue}, {Miller}, {Mahabal}, {Bellm},
  {Patterson}, {Juri{\'c}}, {Golkhou}, {Ofek}, {Walters}, {Graham}, {Kasliwal},
  {Dekany}, {Kupfer}, {Burdge}, {Cannella}, {Barlow}, {Van Sistine}, {Giomi},
  {Fremling}, {Blagorodnova}, {Levitan}, {Riddle}, {Smith}, {Helou}, {Prince},
  \& {Kulkarni}}]{Masci19}
{Masci}, F.~J., {Laher}, R.~R., {Rusholme}, B., {et~al.} 2019, \pasp, 131,
  018003, \dodoi{10.1088/1538-3873/aae8ac}

\bibitem[{{Mej{\'\i}a-Restrepo} {et~al.}(2022){Mej{\'\i}a-Restrepo},
  {Trakhtenbrot}, {Koss}, {Oh}, {den Brok}, {Stern}, {Powell}, {Ricci},
  {Caglar}, {Ricci}, {Bauer}, {Treister}, {Harrison}, {Urry}, {Ananna},
  {Asmus}, {Assef}, {B{\"a}r}, {Bessiere}, {Burtscher}, {Ichikawa}, {Kakkad},
  {Kamraj}, {Mushotzky}, {Privon}, {Rojas}, {Sani}, {Schawinski}, \&
  {Veilleux}}]{Mejia-Restrepo22}
{Mej{\'\i}a-Restrepo}, J.~E., {Trakhtenbrot}, B., {Koss}, M.~J., {et~al.} 2022,
  \apjs, 261, 5, \dodoi{10.3847/1538-4365/ac6602}

\bibitem[{{Richards} {et~al.}(2003){Richards}, {Hall}, {Vanden Berk},
  {Strauss}, {Schneider}, {Weinstein}, {Reichard}, {York}, {Knapp}, {Fan},
  {Ivezi{\'c}}, {Brinkmann}, {Budav{\'a}ri}, {Csabai}, \&
  {Nichol}}]{Richards03}
{Richards}, G.~T., {Hall}, P.~B., {Vanden Berk}, D.~E., {et~al.} 2003, \aj,
  126, 1131, \dodoi{10.1086/377014}

\bibitem[{{Risaliti} {et~al.}(2005){Risaliti}, {Elvis}, {Fabbiano}, {Baldi}, \&
  {Zezas}}]{Risaliti05}
{Risaliti}, G., {Elvis}, M., {Fabbiano}, G., {Baldi}, A., \& {Zezas}, A. 2005,
  \apjl, 623, L93, \dodoi{10.1086/430252}

\bibitem[{{Risaliti} {et~al.}(2002){Risaliti}, {Elvis}, \&
  {Nicastro}}]{Risaliti02}
{Risaliti}, G., {Elvis}, M., \& {Nicastro}, F. 2002, \apj, 571, 234,
  \dodoi{10.1086/324146}

\bibitem[{{Ross} {et~al.}(2020){Ross}, {Graham}, {Calderone}, {Ford},
  {McKernan}, \& {Stern}}]{Ross20}
{Ross}, N.~P., {Graham}, M.~J., {Calderone}, G., {et~al.} 2020, \mnras, 498,
  2339, \dodoi{10.1093/mnras/staa2415}

\bibitem[{{Ruan} {et~al.}(2016){Ruan}, {Anderson}, {Cales}, {Eracleous},
  {Green}, {Morganson}, {Runnoe}, {Shen}, {Wilkinson}, {Blanton}, {Dwelly},
  {Georgakakis}, {Greene}, {LaMassa}, {Merloni}, \& {Schneider}}]{Ruan16}
{Ruan}, J.~J., {Anderson}, S.~F., {Cales}, S.~L., {et~al.} 2016, \apj, 826,
  188, \dodoi{10.3847/0004-637X/826/2/188}

\bibitem[{{Rumbaugh} {et~al.}(2018){Rumbaugh}, {Shen}, {Morganson}, {Liu},
  {Banerji}, {McMahon}, {Abdalla}, {Benoit-L{\'e}vy}, {Bertin}, {Brooks},
  {Buckley-Geer}, {Capozzi}, {Carnero Rosell}, {Carrasco Kind}, {Carretero},
  {Cunha}, {D'Andrea}, {da Costa}, {DePoy}, {Desai}, {Doel}, {Frieman},
  {Garc{\'\i}a-Bellido}, {Gruen}, {Gruendl}, {Gschwend}, {Gutierrez},
  {Honscheid}, {James}, {Kuehn}, {Kuhlmann}, {Kuropatkin}, {Lima}, {Maia},
  {Marshall}, {Martini}, {Menanteau}, {Plazas}, {Reil}, {Roodman}, {Sanchez},
  {Scarpine}, {Schindler}, {Schubnell}, {Sheldon}, {Smith}, {Soares-Santos},
  {Sobreira}, {Suchyta}, {Swanson}, {Walker}, {Wester}, \& {DES
  Collaboration}}]{Rumbaugh18}
{Rumbaugh}, N., {Shen}, Y., {Morganson}, E., {et~al.} 2018, \apj, 854, 160,
  \dodoi{10.3847/1538-4357/aaa9b6}

\bibitem[{{Runnoe} {et~al.}(2016){Runnoe}, {Cales}, {Ruan}, {Eracleous},
  {Anderson}, {Shen}, {Green}, {Morganson}, {LaMassa}, {Greene}, {Dwelly},
  {Schneider}, {Merloni}, {Georgakakis}, \& {Roman-Lopes}}]{Runnoe16}
{Runnoe}, J.~C., {Cales}, S., {Ruan}, J.~J., {et~al.} 2016, \mnras, 455, 1691,
  \dodoi{10.1093/mnras/stv2385}

\bibitem[{{Salim} \& {Narayanan}(2020)}]{SN20}
{Salim}, S., \& {Narayanan}, D. 2020, \araa, 58, 529,
  \dodoi{10.1146/annurev-astro-032620-021933}

\bibitem[{{Scepi} {et~al.}(2021){Scepi}, {Begelman}, \& {Dexter}}]{Scepi21}
{Scepi}, N., {Begelman}, M.~C., \& {Dexter}, J. 2021, \mnras, 502, L50,
  \dodoi{10.1093/mnrasl/slab002}

\bibitem[{{Schlegel} {et~al.}(1998){Schlegel}, {Finkbeiner}, \&
  {Davis}}]{SFD98}
{Schlegel}, D.~J., {Finkbeiner}, D.~P., \& {Davis}, M. 1998, \apj, 500, 525,
  \dodoi{10.1086/305772}

\bibitem[{{Shen}(2021)}]{Shen21}
{Shen}, Y. 2021, \apj, 921, 70, \dodoi{10.3847/1538-4357/ac1ce4}

\bibitem[{{Shen} {et~al.}(2011){Shen}, {Richards}, {Strauss}, {Hall},
  {Schneider}, {Snedden}, {Bizyaev}, {Brewington}, {Malanushenko},
  {Malanushenko}, {Oravetz}, {Pan}, \& {Simmons}}]{Shen11}
{Shen}, Y., {Richards}, G.~T., {Strauss}, M.~A., {et~al.} 2011, \apjs, 194, 45,
  \dodoi{10.1088/0067-0049/194/2/45}

\bibitem[{{Shen} {et~al.}(2019){Shen}, {Hall}, {Horne}, {Zhu}, {McGreer},
  {Simm}, {Trump}, {Kinemuchi}, {Brandt}, {Green}, {Grier}, {Guo}, {Ho},
  {Homayouni}, {Jiang}, {I-Hsiu Li}, {Morganson}, {Petitjean}, {Richards},
  {Schneider}, {Starkey}, {Wang}, {Chambers}, {Kaiser}, {Kudritzki}, {Magnier},
  \& {Waters}}]{Shen19}
{Shen}, Y., {Hall}, P.~B., {Horne}, K., {et~al.} 2019, \apjs, 241, 34,
  \dodoi{10.3847/1538-4365/ab074f}

\bibitem[{{Sheng} {et~al.}(2017){Sheng}, {Wang}, {Jiang}, {Yang}, {Yan}, {Dou},
  \& {Peng}}]{Sheng17}
{Sheng}, Z., {Wang}, T., {Jiang}, N., {et~al.} 2017, \apjl, 846, L7,
  \dodoi{10.3847/2041-8213/aa85de}

\bibitem[{{Smee} {et~al.}(2013){Smee}, {Gunn}, {Uomoto}, {Roe}, {Schlegel},
  {Rockosi}, {Carr}, {Leger}, {Dawson}, {Olmstead}, {Brinkmann}, {Owen},
  {Barkhouser}, {Honscheid}, {Harding}, {Long}, {Lupton}, {Loomis}, {Anderson},
  {Annis}, {Bernardi}, {Bhardwaj}, {Bizyaev}, {Bolton}, {Brewington}, {Briggs},
  {Burles}, {Burns}, {Castander}, {Connolly}, {Davenport}, {Ebelke}, {Epps},
  {Feldman}, {Friedman}, {Frieman}, {Heckman}, {Hull}, {Knapp}, {Lawrence},
  {Loveday}, {Mannery}, {Malanushenko}, {Malanushenko}, {Merrelli}, {Muna},
  {Newman}, {Nichol}, {Oravetz}, {Pan}, {Pope}, {Ricketts}, {Shelden},
  {Sandford}, {Siegmund}, {Simmons}, {Smith}, {Snedden}, {Schneider},
  {SubbaRao}, {Tremonti}, {Waddell}, \& {York}}]{Smee13}
{Smee}, S.~A., {Gunn}, J.~E., {Uomoto}, A., {et~al.} 2013, \aj, 146, 32,
  \dodoi{10.1088/0004-6256/146/2/32}

\bibitem[{{Stern} {et~al.}(2018){Stern}, {McKernan}, {Graham}, {Ford}, {Ross},
  {Meisner}, {Assef}, {Balokovi{\'c}}, {Brightman}, {Dey}, {Drake},
  {Djorgovski}, {Eisenhardt}, \& {Jun}}]{Stern18}
{Stern}, D., {McKernan}, B., {Graham}, M.~J., {et~al.} 2018, \apj, 864, 27,
  \dodoi{10.3847/1538-4357/aac726}

\bibitem[{{Stern} \& {Laor}(2012)}]{SternLaor12}
{Stern}, J., \& {Laor}, A. 2012, \mnras, 423, 600,
  \dodoi{10.1111/j.1365-2966.2012.20901.x}

\bibitem[{{Trakhtenbrot} {et~al.}(2019){Trakhtenbrot}, {Arcavi}, {MacLeod},
  {Ricci}, {Kara}, {Graham}, {Stern}, {Harrison}, {Burke}, {Hiramatsu},
  {Hosseinzadeh}, {Howell}, {Smartt}, {Rest}, {Prieto}, {Shappee}, {Holoien},
  {Bersier}, {Filippenko}, {Brink}, {Zheng}, {Li}, {Remillard}, \&
  {Loewenstein}}]{Trakhtenbrot19}
{Trakhtenbrot}, B., {Arcavi}, I., {MacLeod}, C.~L., {et~al.} 2019, \apj, 883,
  94, \dodoi{10.3847/1538-4357/ab39e4}

\bibitem[{Virtanen {et~al.}(2020)Virtanen, Gommers, Oliphant, Haberland, Reddy,
  Cournapeau, Burovski, Peterson, Weckesser, Bright, {van der Walt}, Brett,
  Wilson, Millman, Mayorov, Nelson, Jones, Kern, Larson, Carey, Polat, Feng,
  Moore, {VanderPlas}, Laxalde, Perktold, Cimrman, Henriksen, Quintero, Harris,
  Archibald, Ribeiro, Pedregosa, {van Mulbregt}, \& {SciPy 1.0
  Contributors}}]{SciPy20}
Virtanen, P., Gommers, R., Oliphant, T.~E., {et~al.} 2020, Nature Methods, 17,
  261, \dodoi{10.1038/s41592-019-0686-2}

\bibitem[{{Wang} {et~al.}(2020){Wang}, {Shen}, {Jiang}, {Grier}, {Horne},
  {Homayouni}, {Peterson}, {Trump}, {Brandt}, {Hall}, {Ho}, {Li}, {Hernandez
  Santisteban}, {Kinemuchi}, {McGreer}, \& {Schneider}}]{Wang20}
{Wang}, S., {Shen}, Y., {Jiang}, L., {et~al.} 2020, \apj, 903, 51,
  \dodoi{10.3847/1538-4357/abb36d}

\bibitem[{{Wright} {et~al.}(2010){Wright}, {Eisenhardt}, {Mainzer}, {Ressler},
  {Cutri}, {Jarrett}, {Kirkpatrick}, {Padgett}, {McMillan}, {Skrutskie},
  {Stanford}, {Cohen}, {Walker}, {Mather}, {Leisawitz}, {Gautier}, {McLean},
  {Benford}, {Lonsdale}, {Blain}, {Mendez}, {Irace}, {Duval}, {Liu}, {Royer},
  {Heinrichsen}, {Howard}, {Shannon}, {Kendall}, {Walsh}, {Larsen}, {Cardon},
  {Schick}, {Schwalm}, {Abid}, {Fabinsky}, {Naes}, \& {Tsai}}]{Wright10}
{Wright}, E.~L., {Eisenhardt}, P. R.~M., {Mainzer}, A.~K., {et~al.} 2010, \aj,
  140, 1868, \dodoi{10.1088/0004-6256/140/6/1868}

\bibitem[{{Xie} {et~al.}(2017){Xie}, {Li}, \& {Hao}}]{Xie17}
{Xie}, Y., {Li}, A., \& {Hao}, L. 2017, \apjs, 228, 6,
  \dodoi{10.3847/1538-4365/228/1/6}

\bibitem[{{Yan} {et~al.}(2019){Yan}, {Wang}, {Jiang}, {Stern}, {Dou},
  {Fremling}, {Graham}, {Drake}, {Yang}, {Burdge}, \& {Kasliwal}}]{Yan19}
{Yan}, L., {Wang}, T., {Jiang}, N., {et~al.} 2019, \apj, 874, 44,
  \dodoi{10.3847/1538-4357/ab074b}

\bibitem[{{Yang} {et~al.}(2018){Yang}, {Wu}, {Fan}, {Jiang}, {McGreer},
  {Shangguan}, {Yao}, {Wang}, {Joshi}, {Green}, {Wang}, {Feng}, {Fu}, {Yang},
  \& {Liu}}]{Yang18}
{Yang}, Q., {Wu}, X.-B., {Fan}, X., {et~al.} 2018, \apj, 862, 109,
  \dodoi{10.3847/1538-4357/aaca3a}

\bibitem[{{Yang} {et~al.}(2020){Yang}, {Shen}, {Liu}, {Aguena}, {Annis},
  {Avila}, {Banerji}, {Bertin}, {Brooks}, {Burke}, {Carnero Rosell}, {Carrasco
  Kind}, {da Costa}, {De Vicente}, {Desai}, {Diehl}, {Doel}, {Flaugher},
  {Fosalba}, {Frieman}, {Garcia-Bellido}, {Gerdes}, {Gruen}, {Gruendl},
  {Gschwend}, {Gutierrez}, {Hinton}, {Hollowood}, {Honscheid}, {Kuropatkin},
  {Maia}, {March}, {Marshall}, {Martini}, {Melchior}, {Menanteau}, {Miquel},
  {Paz-Chinchon}, {Malag{\'o}n}, {Romer}, {Sanchez}, {Scarpine}, {Schubnell},
  {Serrano}, {Sevilla}, {Smith}, {Suchyta}, {Tarle}, {Varga}, \&
  {Wilkinson}}]{Yang20}
{Yang}, Q., {Shen}, Y., {Liu}, X., {et~al.} 2020, \apj, 900, 58,
  \dodoi{10.3847/1538-4357/aba59b}

\bibitem[{{York} {et~al.}(2000){York}, {Adelman}, {Anderson}, {Anderson},
  {Annis}, {Bahcall}, {Bakken}, {Barkhouser}, {Bastian}, {Berman}, {Boroski},
  {Bracker}, {Briegel}, {Briggs}, {Brinkmann}, {Brunner}, {Burles}, {Carey},
  {Carr}, {Castander}, {Chen}, {Colestock}, {Connolly}, {Crocker}, {Csabai},
  {Czarapata}, {Davis}, {Doi}, {Dombeck}, {Eisenstein}, {Ellman}, {Elms},
  {Evans}, {Fan}, {Federwitz}, {Fiscelli}, {Friedman}, {Frieman}, {Fukugita},
  {Gillespie}, {Gunn}, {Gurbani}, {de Haas}, {Haldeman}, {Harris}, {Hayes},
  {Heckman}, {Hennessy}, {Hindsley}, {Holm}, {Holmgren}, {Huang}, {Hull},
  {Husby}, {Ichikawa}, {Ichikawa}, {Ivezi{\'c}}, {Kent}, {Kim}, {Kinney},
  {Klaene}, {Kleinman}, {Kleinman}, {Knapp}, {Korienek}, {Kron}, {Kunszt},
  {Lamb}, {Lee}, {Leger}, {Limmongkol}, {Lindenmeyer}, {Long}, {Loomis},
  {Loveday}, {Lucinio}, {Lupton}, {MacKinnon}, {Mannery}, {Mantsch}, {Margon},
  {McGehee}, {McKay}, {Meiksin}, {Merelli}, {Monet}, {Munn}, {Narayanan},
  {Nash}, {Neilsen}, {Neswold}, {Newberg}, {Nichol}, {Nicinski}, {Nonino},
  {Okada}, {Okamura}, {Ostriker}, {Owen}, {Pauls}, {Peoples}, {Peterson},
  {Petravick}, {Pier}, {Pope}, {Pordes}, {Prosapio}, {Rechenmacher}, {Quinn},
  {Richards}, {Richmond}, {Rivetta}, {Rockosi}, {Ruthmansdorfer}, {Sandford},
  {Schlegel}, {Schneider}, {Sekiguchi}, {Sergey}, {Shimasaku}, {Siegmund},
  {Smee}, {Smith}, {Snedden}, {Stone}, {Stoughton}, {Strauss}, {Stubbs},
  {SubbaRao}, {Szalay}, {Szapudi}, {Szokoly}, {Thakar}, {Tremonti}, {Tucker},
  {Uomoto}, {Vanden Berk}, {Vogeley}, {Waddell}, {Wang}, {Watanabe},
  {Weinberg}, {Yanny}, {Yasuda}, \& {SDSS Collaboration}}]{York2000}
{York}, D.~G., {Adelman}, J., {Anderson}, John~E., J., {et~al.} 2000, \aj, 120,
  1579, \dodoi{10.1086/301513}

\end{thebibliography}
\bibliographystyle{aasjournal}

\end{document}

% End of file `sample63.tex'.